\documentclass[floatfix,twocolumn,showpacs,preprintnumbers,amsmath,amssymb,pra,superscriptaddress,longbibliography]{revtex4-1}
\usepackage{color}
\usepackage[usenames,dvipsnames,svgnames,table]{xcolor}
\usepackage[colorlinks=true,linkcolor=blue,urlcolor=blue,citecolor=blue]{hyperref}
\usepackage{mathtools}
\usepackage{graphicx}
\usepackage{dcolumn}
\usepackage{array}
\usepackage{lipsum}
\usepackage{bm}
\usepackage{amssymb}
\usepackage{multirow}
\usepackage{tabularx}
\usepackage{amsmath}
\usepackage{braket}
\usepackage{csquotes}
\graphicspath{{plots/}}
 \usepackage{lipsum}
\usepackage{mathrsfs}
\usepackage{MnSymbol}
\usepackage{siunitx}
\usepackage{wasysym}
\usepackage{todonotes}

\newcommand{\beq}{\begin{equation}}
\newcommand{\eeq}{\end{equation}}
\newcommand{\bea}{\begin{eqnarray}}
\newcommand{\eea}{\end{eqnarray}}

\newcommand{\qv}{{\bf q}}

\newcommand{\IPD}{{\Delta_\text{IPD}}}

\renewcommand{\vec}[1]{\mathbf{#1}}
\begin{document}

\title{Estimating ionization states and continuum lowering from \emph{ab initio} path integral Monte Carlo simulations for warm dense hydrogen}

\author{Hannah M.~Bellenbaum}
\email{h.bellenbaum@hzdr.de}
\affiliation{Center for Advanced Systems Understanding (CASUS), D-02826 G\"orlitz, Germany}
\affiliation{Helmholtz-Zentrum Dresden-Rossendorf (HZDR), D-01328 Dresden, Germany}
\affiliation{Institut f\"ur Physik, Universit\"at Rostock, D-18057 Rostock, Germany}

\author{Maximilian~P.~B\"ohme}
\affiliation{Lawrence Livermore National Laboratory (LLNL), California 94550 Livermore, USA}

\author{Michael Bonitz}
\affiliation{Institut f\"ur Theoretische Physik und Astrophysik, Christian-Albrechts-Universit\"at zu Kiel, D-24098 Kiel, Germany}


\author{Tilo~D\"oppner}
\affiliation{Lawrence Livermore National Laboratory (LLNL), California 94550 Livermore, USA}

\author{Luke~B.~Fletcher}
\affiliation{SLAC National Accelerator Laboratory, Menlo Park California 94309, USA}

\author{Thomas~Gawne}
\affiliation{Center for Advanced Systems Understanding (CASUS), D-02826 G\"orlitz, Germany}
\affiliation{Helmholtz-Zentrum Dresden-Rossendorf (HZDR), D-01328 Dresden, Germany}

\author{Dominik~Kraus}
\affiliation{Institut f\"ur Physik, Universit\"at Rostock, D-18057 Rostock, Germany}
\affiliation{Helmholtz-Zentrum Dresden-Rossendorf (HZDR), D-01328 Dresden, Germany}

\author{Zhandos~A.~Moldabekov}
\affiliation{Center for Advanced Systems Understanding (CASUS), D-02826 G\"orlitz, Germany}
\affiliation{Helmholtz-Zentrum Dresden-Rossendorf (HZDR), D-01328 Dresden, Germany}

\author{Sebastian Schwalbe}
\affiliation{Center for Advanced Systems Understanding (CASUS), D-02826 G\"orlitz, Germany}
\affiliation{Helmholtz-Zentrum Dresden-Rossendorf (HZDR), D-01328 Dresden, Germany}

\author{Jan Vorberger}
\affiliation{Helmholtz-Zentrum Dresden-Rossendorf (HZDR), D-01328 Dresden, Germany}

\author{Tobias Dornheim}
\email{t.dornheim@hzdr.de}
\affiliation{Center for Advanced Systems Understanding (CASUS), D-02826 G\"orlitz, Germany}
\affiliation{Helmholtz-Zentrum Dresden-Rossendorf (HZDR), D-01328 Dresden, Germany}

\begin{abstract}
Warm dense matter (WDM) is an active field of research, with applications ranging from astrophysics to inertial confinement fusion.
Ionization degree and continuum lowering are important quantities to understand how materials behave under these conditions, but can be difficult to diagnose since experimental campaigns are limited and often require model-dependent analysis.
This is especially true for hydrogen, which has a comparably low scattering cross section, making high quality data particularly difficult to obtain.
Consequently, building equation of state tables often relies on exact simulations in combination with untested approximations to extract properties from experiments.
Here, we investigate an approach for extracting the ionization potential depression and ionization degree -- quantities which are otherwise not directly accessible from the physical model -- from exact ab initio path integral Monte Carlo (PIMC) simulations utilizing a chemical model.
In contrast to experimental measurements, where noise and non-equilibrium effects add to the uncertainty of the inferred parameters, PIMC simulations provide a clean signal with well-defined thermodynamic conditions.
Comparisons against commonly used models show a qualitative agreement, but we find deviations primarily for the high density and high temperature cases.
We also demonstrate the decreasing sensitivity of the dynamic structure factor with respect to both ionization and continuum lowering for increasing scattering angles in x-ray Thomson scattering experiments.
Our work has important implications for the design of future experiments, but also offers qualitative understanding of structure factors and the imaginary-time correlation function obtained from exact quantum Monte Carlo simulations.
\end{abstract}

\maketitle




\section{Introduction\label{sec:introduction}}

Hydrogen is the most abundant element in the universe; it forms a big component of stellar objects and its properties and structures fundamentally dictate the evolution of many astrophysical systems.
As such, it has been studied extensively as a theoretical system~\cite{Bonitz_2024}, but experimental measurements are still hindered e.g.~by the fact that is has a very low scattering cross section~\cite{Fletcher_2016}.
Consequently, developing equation of state (EOS) tables and validating them relies heavily on theoretical models, most commonly the Thomas-Fermi model~\cite{Murillo_2013}, or exact simulations.
The interplay between quantum degeneracy and Coulomb interactions, and the transition between the condensed and plasma phases, however, make the warm dense matter regime very complex and extremely difficult to describe theoretically~\cite{new_POP}.
Experimentally validating any theoretical model is also complicated by the inherent model-dependent nature of analyzing experiments in the high energy density regime~\cite{Dornheim_2022_NatCommun,siegfried_review,Bonitz_2024}.
The direct measurement of plasma parameters like density and temperature is often not possible at these conditions: high temperature plasmas created at research facilities usually only exist on very short timescales and the extreme conditions typically destroy a sample~\cite{Landen_2024}, meaning most diagnostics rely on some indirect measurement containing inherent model assumptions~\cite{siegfried_review,Dornheim_2022_NatCommun}.

While exact methods to describe hydrogen in the warm dense matter regime are scarce,  recent developments~\cite{Dornheim_JCP_xi_2023,Dornheim_JPCL_2024,Dornheim_Science_2024} in \emph{ab initio} path integral Monte Carlo (PIMC)~\cite{cep} simulations allow for the calculation of the Laplace transform of the dynamic structure factor for light elements such as hydrogen \cite{Dornheim_MRE_2024}.
These simulations are limited in system size and with respect to the accessible densities and temperatures, however, due to the fermion sign problem \cite{Bonitz_2024,new_POP,dornheim_sign_problem,troyer}, and are computationally expensive to run.
Nevertheless, PIMC and other quantum mechanical modeling tools like density functional theory (DFT) are often used in constructing EOS tables (e.g.~\cite{Militzer_PRE_2021,LA_McHardy_2018}).
These constitute an indispensable input for the description of material behavior under WDM conditions in large scale simulations, like multi-dimensional simulations containing both radiation physics and magneto-hydrodynamic effects that are in turn used to design and optimize ICF experiments, and which have been vital in recent developments to reach a burning capsule design on the NIF~\cite{Zylstra2022,Kritcher_2022,Marinak_2024,Hurricane_2023}.

Two important, related properties of a plasma are its ionization state and the process of continuum lowering. Continuum lowering (CL) is a fundamental process whereby electrostatic interactions of an ion's electrons with the surrounding plasma lowers the binding energy of the electrons~\cite{griem_1997, Ecker_1963, Stewart_1966}.
An important consequence of CL is that if it exceeds the binding energy of an electron, that electron becomes pressure ionized. CL therefore has a direct impact on the ionization state, EOS, opacity, and transport properties in the plasma.
For many applications CL is treated with approximate numerical and analytical models of ionization potential depression (IPD)~\cite{Ecker_1963, Stewart_1966}.
Direct measurement of CL is challenging since it requires that the plasma's conditions (e.g. temperature, density, and charge states of the ions) are well-known, and that the true continuum edge can be clearly distinguished.
Emission spectroscopy in particular has proved very useful in drawing numerous important conclusions on the nature of CL~\cite{Ciricosta_2012,Hoarty2013,hansen2017changes,hu2022probing}, and has even been used to make direct measurements of CL in materials such as magnesium and aluminum~\cite{Ciricosta2016-nx}.
However, the interpretation of the results has been previously debated and is potentially model-dependent~\cite{Iglesias_2013_IPD, Iglesias_2014_IPD,Ciricosta_2016_Detailed}.
Furthermore, even the direct measurements have seen a recent re-interpretation that lead to different conclusions on the strength of CL in dense plasmas~\cite{Gawne_2023}.
The ionization state of the system, on the other hand, is a quantity that that does not have a clear definition, but is related to the localization of electrons within the system.
A common interpretation of ionization is based on the charge state distribution of the ions -- i.e. counting the number of electrons freed from the ions -- however such a definition is in tension with reality since it (a) requires one to distinguish electrons as either fully bound to ions or completely free, which is generally not the case, and (b) such a definition of ionization appears to underpredict the actual ionization state as inferred from experimental measurements.
Point (a) also raises issues for CL since if electrons are neither fully bound or fully free, then the concept of a ``true'' edge may no longer be possible.

The importance of CL and ionization to studying and making predictions about high energy density systems has lead to numerous experimental and theoretical efforts to understand these properties.
From a theoretical perspective, \textit{ab-initio} methods such as DFT and PIMC seem ideal candidates to understand these properties since they function within the physical picture (i.e. electrons are not distinguished as bound or free), and ionization and CL are inherently present.
However, this also means that these approaches do not give direct access to CL and ionization for the same reason as mentioned above.
Therefore, definitions of these quantities need to be externally imposed on the system, for example by making some approximation on the boundness of the electrons.
A common approach is to say that we can distinguish electrons as either being fully bound to an ion or fully free (the so-called ``chemical model'').
For obtaining an ionization from DFT, there have been proposed definitions based on the free-free Kubo-Greenwood conductivity~\cite{Bethkenhagen_2020} and from counting the continuum electrons~\cite{Gawne_2024}. In both definitions, the free electrons are defined as those in the continuum, but the methods lead to notably different ionization states with better or worse agreement with different experiments~\cite{Tilo_Nature_2023,Whang_1972_Optical,Kobayasi_1972}, which points to the difficulty in pinning down a precise definition even within the chemical model.
Bonitz \emph{et al.}~\cite{Bonitz_2024} have suggested that one can use the Saha equation~\cite{Saha_prs_1921} to extract an estimate of the ionization state from PIMC simulations of warm dense hydrogen. This concept was explored in more detail in Ref.~\cite{Bonitz_2025} to further extract estimates for the IPD.
There have also been attempts to extract CL from DFT calculations, for example one method based on the difference in free energy between two frozen-core calculations~\cite{vinko2014density}, and a more recent approach based on the characterizing the spatial localization of electrons around ions~\cite{Gawne_2023}. Although there are similarities in the results, differences emerge due to the how electrons are classed as bound or free.
Finally, we mention very recent work by Zeng \emph{et al.}~\cite{Zeng_PRE_2025} that has extended the chemical model to very dense plasmas with $\rho\gtrsim10\,$g/cc.


Here, we propose a different route motivated by the Chihara decomposition that is most commonly used to interpret x-ray Thomson scattering (XRTS) spectra~\cite{siegfried_review}, and has been previously used to obtain estimates for the degree of ionization~\cite{Kraus2016,Tilo_Nature_2023} and continuum lowering~\cite{kraus_xrts,Fletcher_2014}.
The Chihara decomposition gives one access to different components of a scattering signal since bound and free electrons are treated separately.
Therefore,their individual contributions to the scattering signal can be used to diagnose parameters like ionization or the effect of continuum lowering on the binding energy of a species.
This naturally introduces a definition for the ionization state of a species as the mean number of free electrons per nucleus (i.e. the mean charge state of the ions), by explicitly considering contributions of both to the total scattering signal.
Continuum lowering, or ionization potential depression (IPD), consequently is the effective reduction in the binding energy of a given ion species, due to exchange effects and particle interactions~\cite{Crowley_2014_ipd,Bonitz_2024}.
Thus, by separating scattering contributions of free and bound electrons to calculate a dynamic structure (DSF), we can apply chemical concepts like the ionization to a physical description obtained from PIMC datasets.
Similar to the approach widely used in analyzing XRTS experiments, where synthetic structure factors for a variety of conditions (for example temperature, density, ionization) are forward-fitted to an experimentally measured spectrum to extract best fit parameters~\cite{Kasim_2019}, we propose fitting a chemical DSF to PIMC simulations.
This removes many of the shortcomings of the conventional approach of extracting plasma parameters from an experimental spectrum, mainly the uncertainty involved in the measured signal being a convolution of the DSF and the source-instrument function, experimental noise and possible non-equilibrium and inhomogeneity effects~\cite{Vorberger_PRX_2023}.

Recent developments in extracting model-independent temperature measurements from XRTS using the imaginary-time correlation function (ITCF) approach (see e.g.~\cite{Dornheim_2022_NatCommun,Dornheim_T_follow_up,Bellenbaum_2025}) have removed some of these uncertainties, but measuring density or even ionization still requires forward-modeling.
Moreover, by making use of PIMC simulations for a known density and temperature, we can optimize our approach to only fit for two free parameters: the IPD and the ionization state.
Additionally, we are not limited to a single, or multiple scattering angles, as is often the case in XRTS experiments.
While simultaneous measurements from forward and backward scattering angles have been analyzed (see e.g.~\cite{Bellenbaum_2025}), an experimental XRTS setup is still very much constrained by chamber size, geometry and other diagnostic requirements.
By accessing PIMC simulations over an extended range of scattering wave vectors, corresponding to all accessible scattering angles in an XRTS set-up, we can test the dependence of different models and the sensitivity of extracted parameters to the scattering geometry.
Further, because the Chihara decomposition as a chemical model relies fundamentally on the assumption that free and bound electrons can be clearly separated in their scattering contributions, we can test this approximation under WDM conditions against exact data and its consistency over all scattering angles.
Finally, applying a chemical model to PIMC data gives us qualitative insights into the physics contained within the ITCF and a better understanding of structure factors obtained from PIMC by approximately inverting the Laplace transform.


This paper is organized as follows: Sec. \ref{sec:theory} will go into detail on the theoretical description of the dynamic structure factors in both PIMC and the Chihara decomposition.
Sec. \ref{sec:method} will give details on the simulations and the fitting procedure used, with results given in Sec. \ref{sec:results}.
We finish by summarizing and discussing the results in Sec. \ref{sec:summary}.


\section{Theory\label{sec:theory}}

Outputs from PIMC are given as the imaginary-time correlation function (ITCF)
\begin{eqnarray}\label{eqn:laplace-itcf}
    F_{ee}(\mathbf{q},\tau) = \int_{-\infty}^\infty \textnormal{d}\omega\ S_{ee}(\mathbf{q},\omega) e^{-\tau \hbar\omega}
    \,,
\end{eqnarray}
which is simply given by the two-sided Laplace transform of the full electron-electron dynamic structure factor $S_{ee}$~\cite{Dornheim_MRE_2023,Dornheim_review} in imaginary time $\tau$, where $\omega$ is the energy and $\hbar$ is the reduced Planck constant.
The Chihara decomposition \cite{Chihara_1987} is then used to enforce a definition of ionization state and IPD.
In XRTS experiments, one often distinguishes between inelastic and elastic scattering components~\cite{siegfried_review,Dornheim_2024_preprint}, which can also be applied to the PIMC simulations here:
\begin{eqnarray}
    S_{ee}(\mathbf{q},E) &=& W_R(\mathbf{q}) \delta(E) + S_\textnormal{inel}(\mathbf{q},E), \\ 
    \Rightarrow F_{ee}(\mathbf{q},\tau)  &=& W_R(\mathbf{q}) + F_\textnormal{inel}(\mathbf{q},\tau)
    \,.
\end{eqnarray}
The term $W_R(\qv)$, known as the Rayleigh weight, describes the elastic component, i.e. the effectively static contribution from electrons closely following the ionic motion, and is therefore obtained from PIMC as~\cite{Dornheim_2024_preprint}
\begin{eqnarray}\label{eq:Rayleigh_weight}
    W_R(\mathbf{q}) =  \frac{1}{Z} \rho^2(\mathbf{q}) S_{ii}(\mathbf{q}) = \frac{S^2_{ei}(\mathbf{q})}{S_{ii}(\mathbf{q})}
    \,,
\end{eqnarray}
for the ionization state $Z$, with the definition of the generalized form factor given in terms of the electron-ion and ion-ion contributions $S_{ei}$ and $S_{ei}$ as~\cite{Vorberger_PRE_2015} 
\begin{eqnarray}
    \rho(\mathbf{q}) = f(\mathbf{q}) + q(\mathbf{q}) = \frac{1}{\sqrt{Z_\text{tot}}} \frac{S_{ei}(\mathbf{q})}{S_{ii}(\mathbf{q})}
    \,,
\end{eqnarray}
where $Z_\text{tot}$ is the total charge, $q(\mathbf{q})$ is the screening cloud and $f(\mathbf{q})$ is the form factor.

In the chemical picture, for a multi-species system, by considering the hydrogen as either fully ionized or unionized, and defining the mean ionization state $Z$ in terms of the partial densities of each, the Rayleigh weight is defined as~\cite{Wunsch_2011}:
\begin{equation}
    W_R(q) = \sum_{a,b} \sqrt{x_a x_b} [f_a(\qv) + q_a(\qv)] [f_b(\qv) + q_b(\qv)] S_{ab}(\qv)
    \,,
\end{equation}
for the screening cloud and form factors of species $a$ and $b$ respectively, with partial densities $x_a + x_b=1$.

The inelastic component in the Chihara decomposition contains contributions of both the free and bound electrons:
\begin{eqnarray}
    S_{\text{inel}}(\qv,\omega) &=& S^{\text{ff}}(\qv,\omega) + S^{\text{bf}}(\qv,\omega) \\\nonumber
    &=& \sum_a Z_a^f n_i S_{ee}^{\text{EG}}(\qv,\omega) \\
    &&+ \sum_a Z_a^b n_i S_{ea}^{\text{core}}(\qv,\omega)
    \,.
\end{eqnarray}
The bound-free component $S^\text{bf}$ here contains scattering contributions from bound electrons which are excited to the continuum by Raman transitions, as well as the reverse free-bound process to enforce detailed balance~\cite{boehme2023evidence}.
Compton scattering of the free electrons is taken into account in the $S^\text{ff}$ component.
Here, $n_i$ is the ion number density, $Z_a^f$ is the number of free electrons per nucleus of species $a$ and $Z_a^b$ correspondingly is the number of bound electrons per nucleus.
$S_{ee}^{\text{EG}}$ is the electron-gas structure factor describing the free electrons, whereas $S_{ea}^{\text{core}}$ describes the scattering behavior of the bound electrons scaled using the fraction of bound electrons.
Both are summed over each species $a$ present.

Thus, the mean charge state $Z$ here is directly related to the separation of free and bound contributions in the Chihara decomposition and forms the basis of our definition.
Similarly, ionization potential depression can be defined an approximate description of the reduction of the binding energy $E_b$ of electrons in a dense plasma due to interactions with neighboring particles, as well as exchange effects \cite{Crowley_2014_ipd,Bonitz_2025}, giving an effective binding energy $E_b^\text{eff}$.
The concept takes into account screening of the nuclear charge by the surrounding electrons and ions, which leads to:
\begin{equation}
    E_b^\text{eff} = E_b - \Delta_\text{IPD}
    \,.
\end{equation}
We apply the frequently used simplified relationship here between the scattering angle $\phi$ and the wave vector~\cite{siegfried_review} to relate wave numbers to a typical scattering geometry often applied in XRTS experiments:
\begin{equation}
    \vert \vec{q} \vert = q \approx 2 q_0 \sin(\phi/2)
    \,, \label{eq:delta-ipd}
\end{equation}
where the prefactor $q_0$ is defined as $q_0=\hbar c/2 \omega_0$ for a given beam energy $\omega_0$.
For details on the approximate form and regimes where it starts to become inaccurate, see Ref.~\cite{Dornheim_T_follow_up}.

In the WDM regime, it is often useful to introduce non-dimensional parameters like $r_s$ for the quantum coupling parameter and $\theta$ for the non-dimensional temperature; in particular $\theta=\text{const.}$ defines lines of constant quantum degeneracy.
These are defined in terms of the electron number density $n_e$ and the Bohr radius $a_B$ as \cite{Dornheim_T_follow_up}
\begin{equation}
    r_s = \left(\frac{1}{4 \pi n_e a_B^3} \right)^{1/3}
    \,,
\end{equation}
and
\begin{equation}
    \theta = \frac{k_B T}{E_F} = \frac{2 m_e k_B T }{\hbar^2 q_F^2} = k_B T \frac{2 m_e}{\hbar} \left(3 \pi^2 n_e \right)^{-2/3}
    \,,
\end{equation}
where $m_e$ is the electron mass and $k_B$ is the Boltzmann constant.
The Fermi wave number is defined as $q_F = \left(3 \pi n_e \right)^{1/3}$ and the Fermi energy is given by $E_F = \hbar^2 q_F^2 / (2 m_e)$.

\begin{figure}
    \centering
    \includegraphics[width=0.95\linewidth]{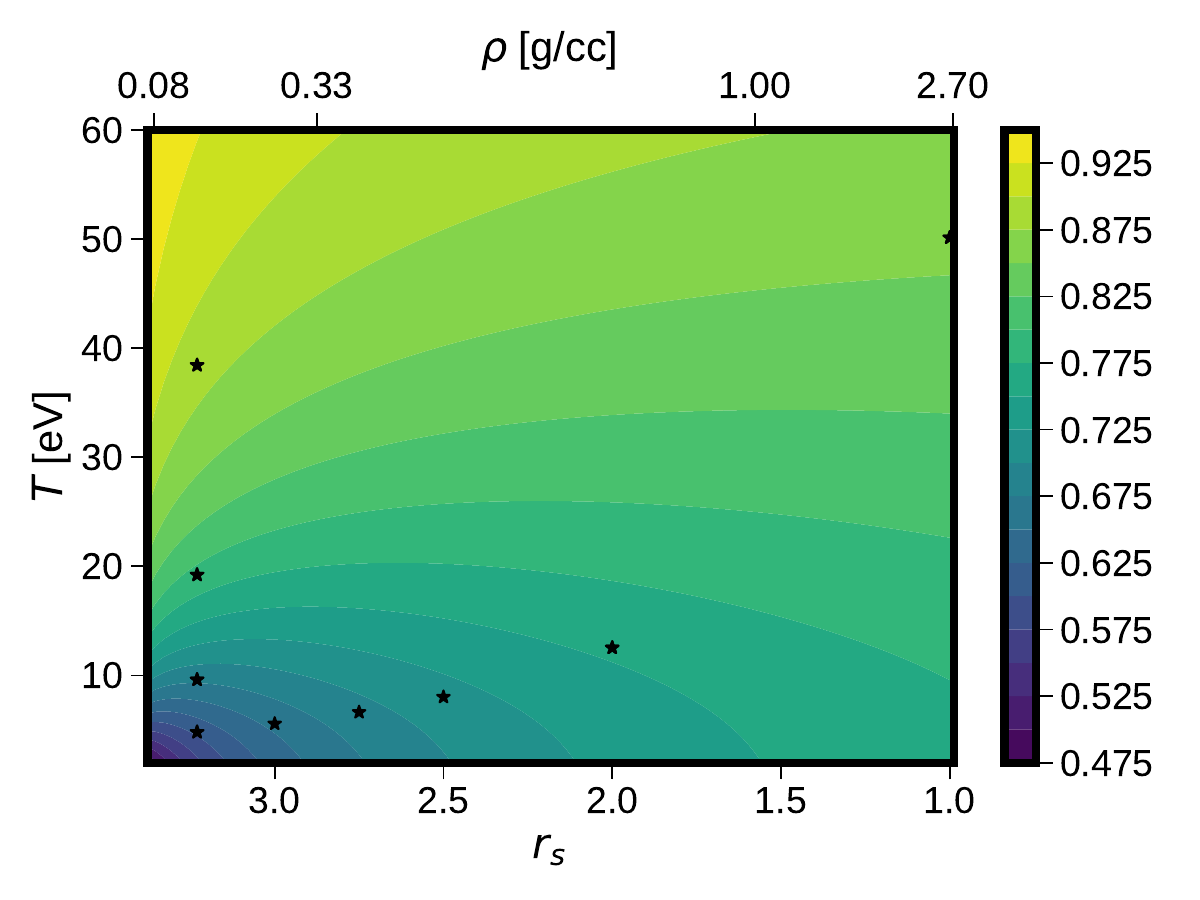}
    \caption{
    Heat map of the Thomas-Fermi estimate of Hydrogen's ionization degree obtained using Murrilo \emph{et al.} implementation~\cite{Murillo_2013} of More \emph{et al.}~\cite{More_1982}.
    Available PIMC simulations are indicated in parameter space by the stars.
    A direct comparison of the estimated ionization states to the Thomas-Fermi model is given in Fig.~\ref{fig:fit-comparison} and~\ref{fig:fit-comparison-const-theta}.
    }
    \label{fig:tf-hydrogen}
\end{figure}

\section{Method}\label{sec:method}

Similarly to the analysis of XRTS experiments, where DSFs convolved with a source-instrument function are fitted against experimentally measured spectra, here we perform such a forward-fitting procedure to PIMC simulations.
Our aim is to extract estimates for both ionization and IPD over all scattering vectors and to test the consistency of the widely used Chihara model.

\subsection{PIMC simulations}\label{subsec:pimc-sims}

The \emph{ab initio} PIMC method~\cite{cep} is based on the celebrated classical isomorphism~\cite{Chandler_JCP_1981} that allows one to map the interacting quantum many-body system of interest onto an effectively classical ensemble of interacting ring polymers.
As a starting point, we consider the canonical partition function (i.e., particle number $N$, volume $\Omega=L^3$ and inverse temperature $\beta=1/k_\textnormal{B}T$ are fixed) in coordinate representation
\begin{eqnarray}\label{eq:Z}
    \Sigma = \sum_{\sigma^\uparrow}\sum_{\sigma^\downarrow} \textnormal{sgn}\left(\sigma^\uparrow,\sigma^\downarrow\right) \int_\Omega \textnormal{d}\mathbf{R}\ \bra{\mathbf{R}} e^{-\beta\hat{H}} \ket{\hat{\pi}_{\sigma^\uparrow,\sigma^\downarrow}\mathbf{R}}\ ,
\end{eqnarray}
where the vector $\mathbf{R}$ contains the coordinates of both electrons and protons. Note that we have to explicitly sum over all permutations $\sigma^\uparrow$ and $\sigma^\downarrow$ of the spin-up and spin-down electrons that are realized by the corresponding exchange operator $\hat{\pi}_{\sigma^\uparrow,\sigma^\downarrow}$ and taken into account with the appropriate sign $\textnormal{sgn}\left(\sigma^\uparrow,\sigma^\downarrow\right)$, which is positive (negative) for an even (odd) number of pair permutations. 
This leads to the notorious fermion cancellation problem~\cite{dornheim_sign_problem,troyer}, which we mitigate by a) averaging over a large number of independent Monte Carlo samples and b) utilizing the recent $\xi$-extrapolation method~\cite{Xiong_JCP_2022,Dornheim_JCP_xi_2023,Dornheim_MRE_2023,Dornheim_Science_2024} at selected parameters.
The key problem that precludes the straightforward evaluation of Eq.~(\ref{eq:Z}) is caused by the matrix elements of the density operator $\hat{\rho}=e^{-\beta\hat{H}}$ since the potential and kinetic contributions to the full Hamiltonian $\hat{H}=\hat{V}+\hat{K}$ do not commute. To circumvent this obstacle, we use the usual exact semi-group property of $\hat{\rho}$ (i.e., introducing $P$ factors at a reduced inverse temperature $\epsilon=\beta/P$) and combine it with the primitive factorization $e^{-\epsilon\hat{H}}\approx e^{-\epsilon\hat{V}}e^{-\epsilon\hat{K}}$ that becomes exact in the limit of large $P$. 
In this way, the original partition function [Eq.~(\ref{eq:Z})] is being expressed as the integral over all possible paths of particle coordinates along the imaginary-time $t=-i\hbar\tau$, with $\tau\in[0,\beta]$, which are generated stochastically using modern implementations~\cite{boninsegni1,boninsegni2,Dornheim_PRB_nk_2021} of the Metropolis algorithm~\cite{metropolis}.
Although it is computationally rather expensive, the PIMC method is capable of giving quasi-exact results (i.e., exact within the given Monte Carlo error bars for a given system size $N$) for any physical observable, often making it the method of choice for the computation of input data for other processes such as the exchange--correlation function for thermal density functional theory~\cite{groth_prl,review,status}.
In the context of the present work, we highlight the capability of PIMC to estimate the imaginary-time density--density correlation function
\begin{eqnarray}
    F_{ee}(\mathbf{q},\tau) = \braket{ \hat{n}_e(\mathbf{q},\tau) \hat{n}_e(-\mathbf{q},0) }
\end{eqnarray}
by correlating electronic densities on separate imaginary time slices~\cite{Dornheim_JCP_ITCF_2021}.

All simulations have been performed using the \texttt{ISHTAR} code~\cite{ISHTAR}, which is a canonical adaption~\cite{Dornheim_PRB_nk_2021} of the worm algorithm by Boninsegni \textit{et al.}~\cite{boninsegni1,boninsegni2}. Specifically, we combine a primitive factorization of the Coulomb repulsion between equal charges with the exact pair density matrix between electrons and protons to remove the Coulomb divergence as it has been described in detail in Ref.~\cite{Bohme_PRE_2023}. We use $P=200$ imaginary-time propagators throughout, and the convergence with $P$ has been carefully checked, see also the Supplemental Material of Ref.~\cite{Bohme_PRL_2022}.

\subsection{Extracting fit parameters}\label{subsec:fitting}

\begin{figure}
\centering
\includegraphics[width=0.99\linewidth]{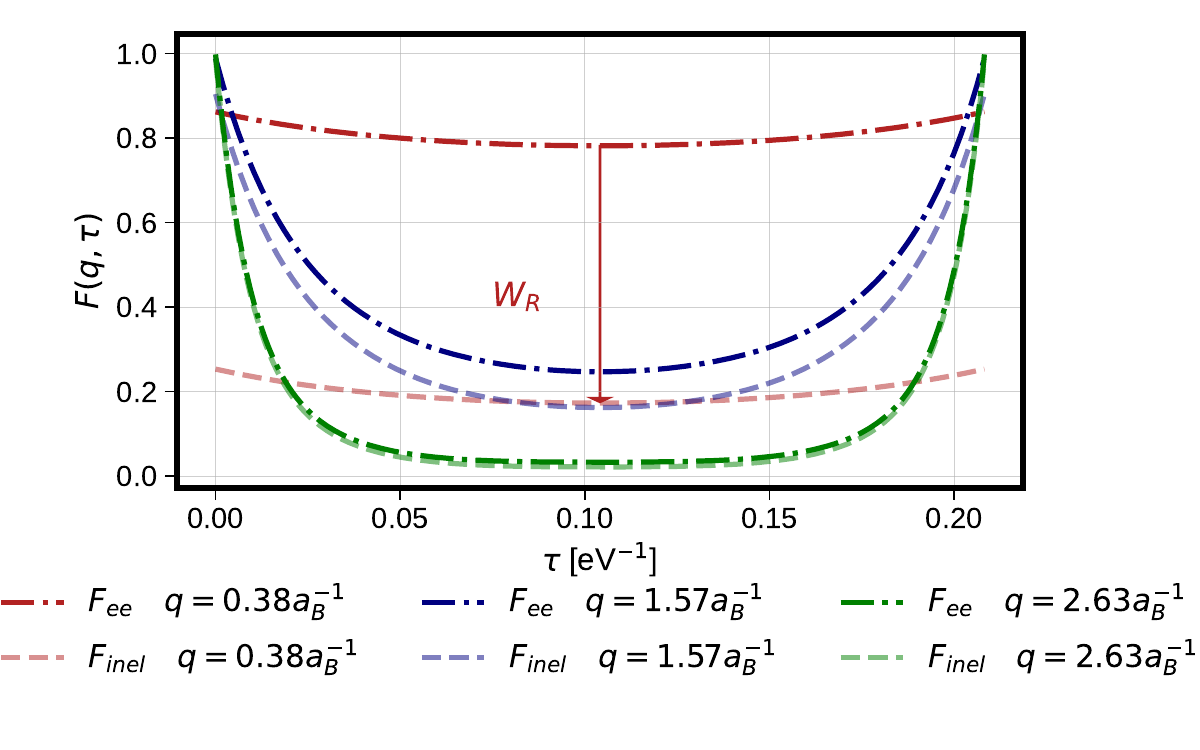}
\caption{\label{fig:ITCF_example}
ITCF for selected $q$ for $N=32$, $r_s=2$, and $\theta=1$ plotted against imaginary time $\tau$.
Here, we plotted the ITCF different scattering wave numbers $q$ to show contributions of the elastic component, also known as the Rayleigh weight $W_R$.
The full ITCF $F_{ee}$ is plotted in the dashdot lines, the inelastic component $F_\text{inel}$ is plotted in the dashed lines and the subtraction of the Rayleigh weight $W_R$ is indicated for the smallest $q$-value using the arrow.
}
\label{fig:hydrogen-itcf}
\end{figure}

Since IPD and ionization degree are inherently related parameters, simultaneously fitting for both might introduce dependencies and overfitting.
Instead, we fit for the ionization degree using the elastic part only and, using those outputs, fit for an IPD using the inelastic component.
This allows us to effectively decouple the two, without having to rely on additional assumptions like the Stewart-Pyatt IPD model.

In Fig.~\ref{fig:hydrogen-itcf} we show PIMC results for $F_{ee}$ for Hydrogen at $N=32$, $r_s=2$, and $\theta=1$ for three different $q$-values.
The ITCF in itself contains a myriad of information on the properties of a material~\cite{Dornheim_MRE_2023,Dornheim_PTR_2023}.
For example, one can directly extract the temperature by considering the detailed balance relation which manifests itself as a symmetry relation in $\tau$~\cite{Dornheim_2022_NatCommun}.
Thus, the minimum always corresponds to the inverse temperature $\beta=1/T=2\tau_\text{min}$.
In addition, the slope of the ITCF around $\tau=0$ gives access to the f-sum rule~\cite{Dornheim_MRE_2023,dornheim2023xray}.
Here, by separating out the elastic and inelastic components of the dynamic structure factor from the available PIMC data, we can independently fit for each to obtain estimates for both the ionization state and the IPD.
Specifically, the inelastic component of the ITCF is obtained from PIMC by simply subtracting the Rayleigh weight from the full ITCF, which is a constant offset in the imaginary-time perspective.
Notably, the contribution of the elastic scattering reduces for increasing scattering wave numbers.

We use the \emph{multi-component scattering spectra} (MCSS) code~\cite{mcss_user_manual} to calculate the synthetic dynamic structure factors of the chemical model, for both the inelastic and elastic components of the Chihara decomposition.
The free-free component is modeled using a numerical fit to the random phase approximation (RPA), with an analytic fit to the effective static local field correction developed in \cite{Dornheim_PRB_ESA_2021,Dornheim_PRB_2021}.
An impulse approximation is used to model the bound-free DSF \cite{Schumacher_1975}, and the reverse process---previously neglected free-bound transitions~\cite{boehme2023evidence}---is included by automatically enforcing the detailed balance relation $S(\mathbf{q},-\omega)=S(\mathbf{q},\omega)e^{-\hbar\omega/k_\textnormal{B}T}$~\cite{quantum_theory}.
Ion-ion and electron-ion interactions are approximated using the Debye-H\"uckel pseudo-potential \cite{Hückel_1924}.

Fit results are obtained by minimizing the mean difference $\delta$ between the PIMC results and the chemical model (CM) defined separately as
\begin{equation}
    \delta_\text{el}= \lvert W_R^{\text{PIMC}} (Z)  - W_R^{\text{CM}} \rvert
    \,,
\end{equation}
for the elastic component to obtain a `best fit` ionization degree $Z_\text{fit}$, and subsequently
\begin{equation}
    \delta_\text{inel}= \frac{1}{P} \sum_{\tau} \big[ F_{\text{PIMC}}(q, \tau) - F_{\text{CM}}(q, \tau) \vert_{Z_\text{fit},\IPD} \big]^2
    \,,
\end{equation}
to obtain a `best fit` estimate for the IPD as $\Delta^\text{fit}_\text{IPD}$.
Hence, we are summing over all $P$ points in $\tau$ for the inelastic component and using the $Z^\text{fit} = \min_Z [\delta_\text{el}]$ estimated from the inelastic component, obtain the IPD from the inelastic one as $\Delta^\text{fit}_\text{IPD} = \min_\IPD [\delta_\text{inel}]$.

Since the Chihara decomposition as an approximate model is quick to run, determining the $\delta$ parameter over the entire plane is very inexpensive and easily done even for a large dataset, especially compared to \emph{ab initio} tools like DFT-MD.
Creating surface plots of this fitting error gives one immediate access to a `best-fit` result as will be shown below, and further allows us to investigate the consistency of our chemical model.

\section{Results\label{sec:results}}

In this section, we present the results from fitting a chemical model to exact PIMC data for warm dense hydrogen to extract estimates for the ionization state and the IPD.
Results are given over a large parameter space in $r_s-\theta$ and over an extended $q$-range.
This allows us to estimate best fit values and test the consistency of the Chihara decomposition for different scattering angles in a typical XRTS experiment.

\begin{figure*}
    \includegraphics[width=0.49\linewidth]{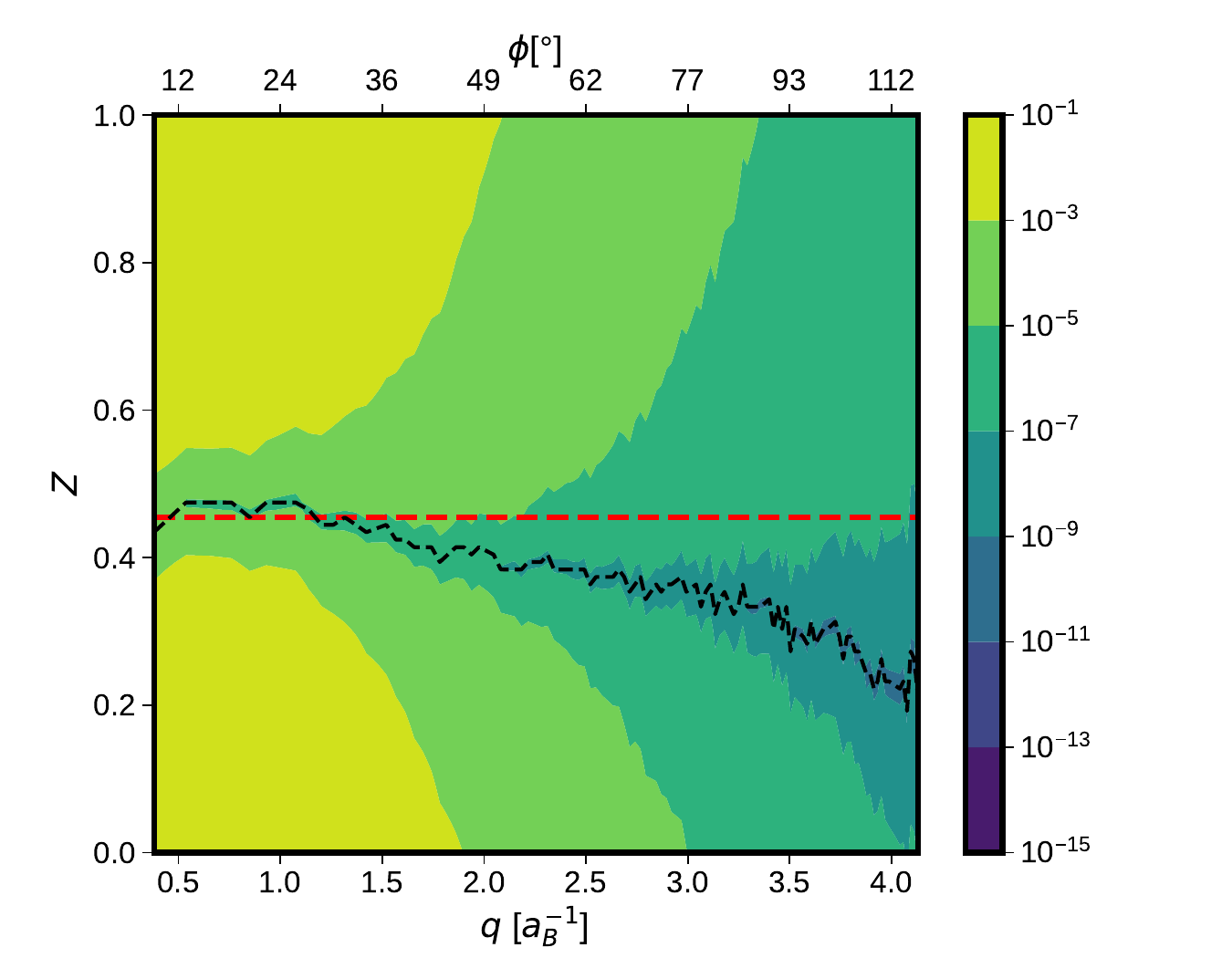}
    \includegraphics[width=0.49\linewidth]{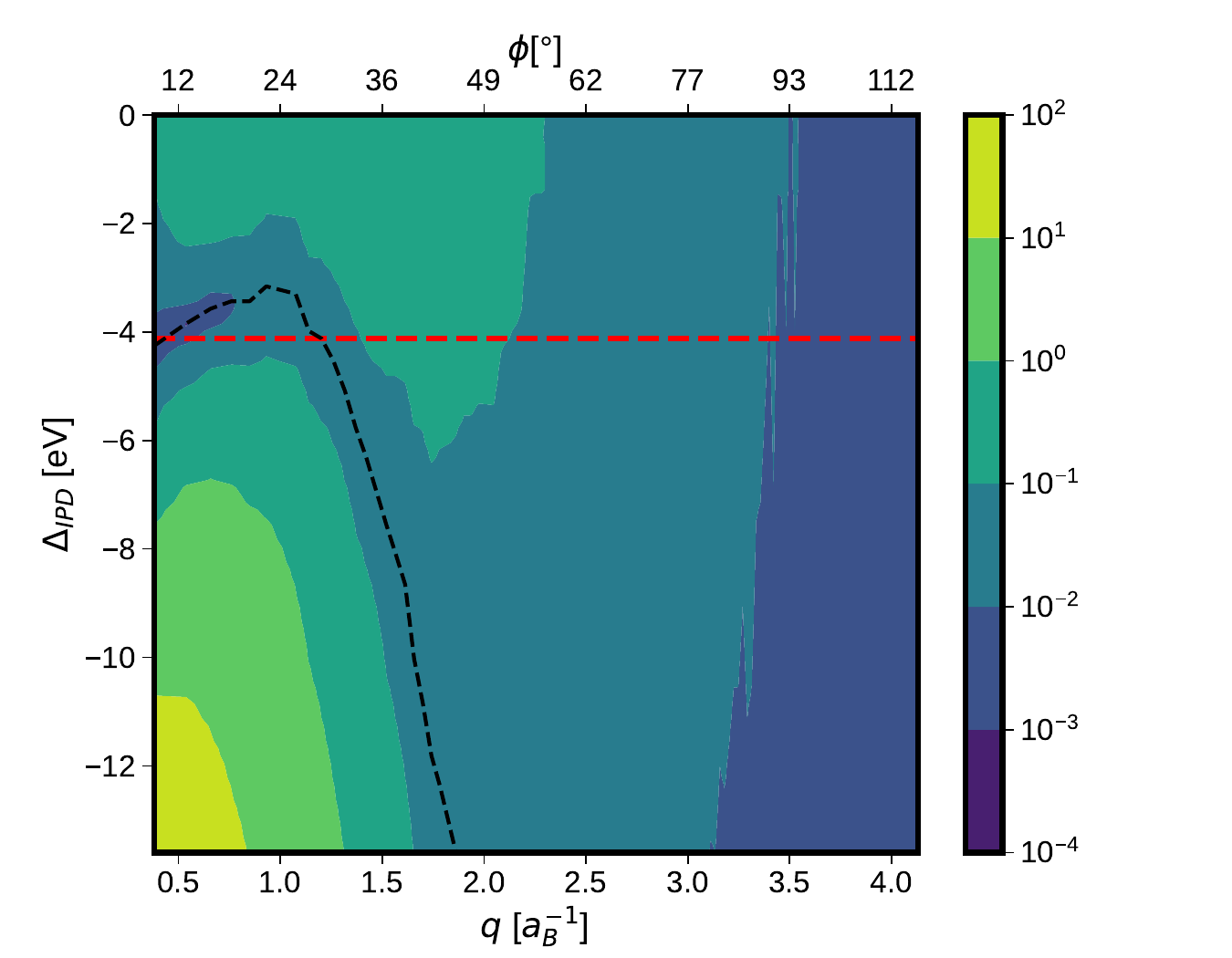}
    \includegraphics[width=0.49\linewidth]{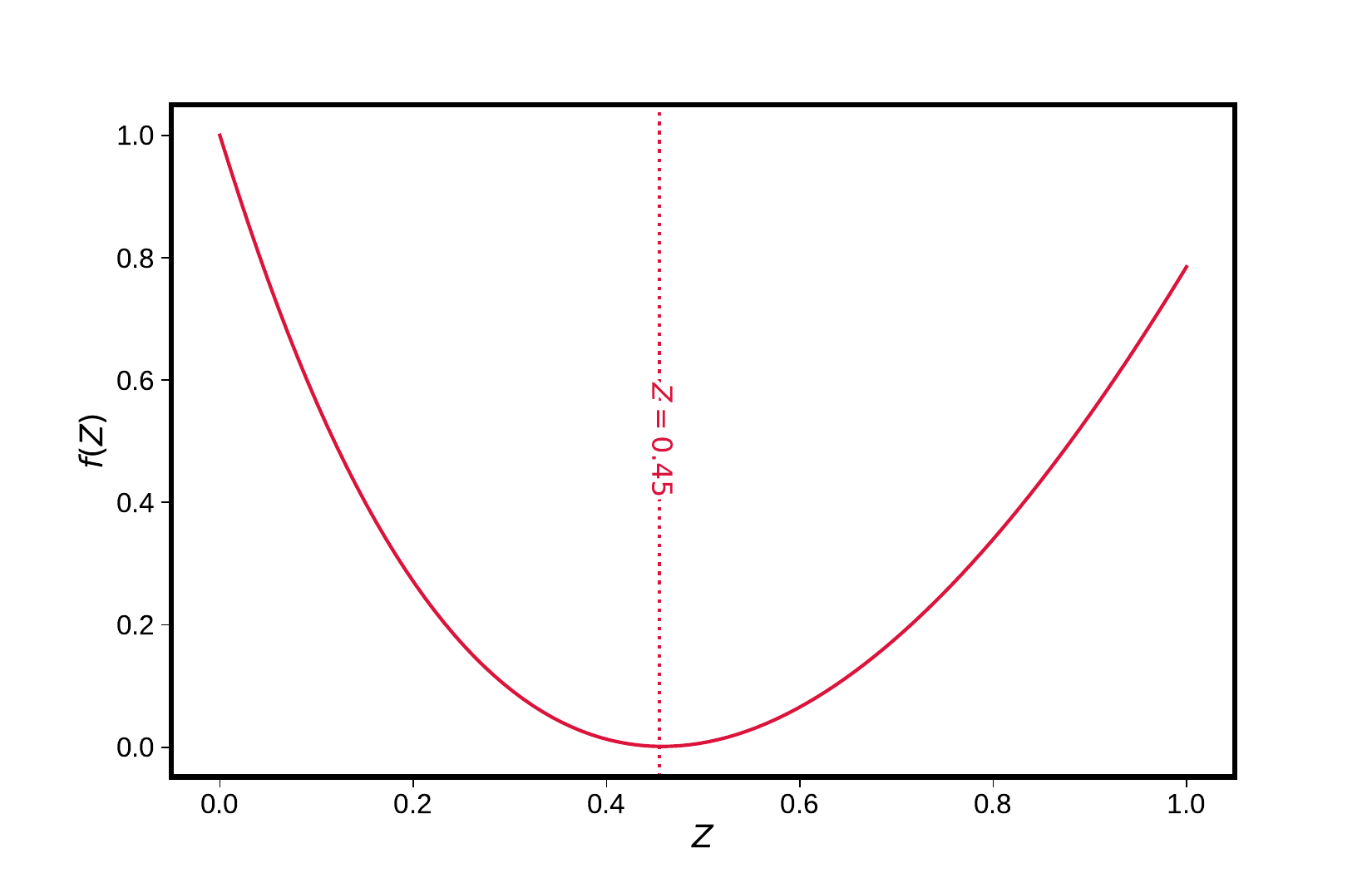}
    \includegraphics[width=0.49\linewidth]{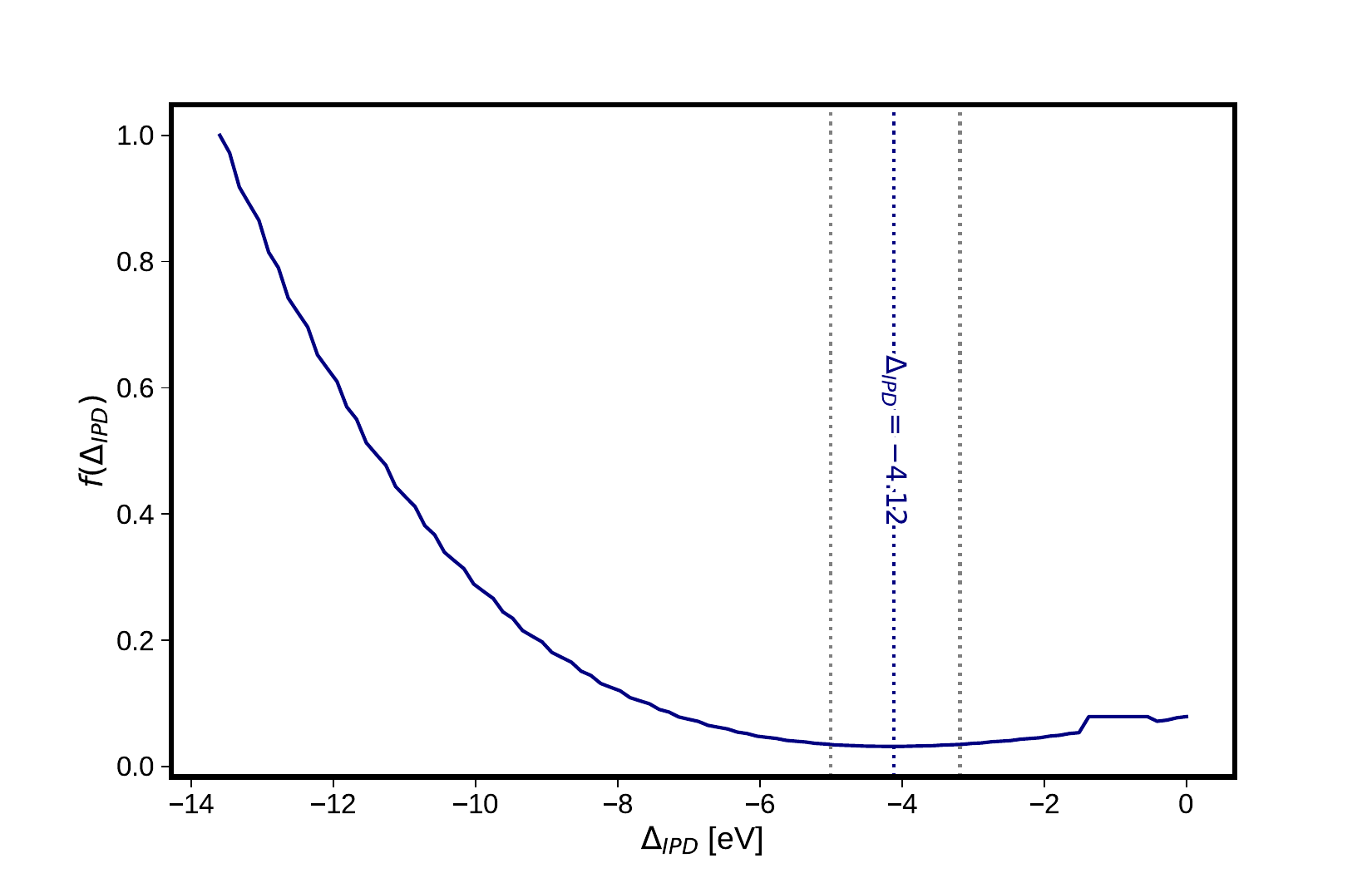}
\caption{Results for the $N=32$, $r_s=3.23$, $\theta=1$ dataset.
Top left: $\delta$-error plot for the Rayleigh weight over $q-Z$ parameter space.
Top right: $\delta$-error plot for the inelastic component over $q-\IPD$ parameter space.
A corresponding scattering angle for a typical beam energy of $\omega_0=9\unit{keV}$ is plotted on the top axis for each heatmap.
The global minimum over all $q$ is indicated by the dashed red line.
Bottom left: The $\delta$-error summed over all points in $q$ for the ionization surface plot.
Bottom right: The $\delta$-error summed over all points in $q$ for the IPD surface plot.
The $q$-dependent minimum is indicated by the dashed black line for both $f(x)$ curves to extract the best-fit result.
Both are arbitrarily normalized to enforce $f(0)=1$ and dotted black lines are used to indicate a 10\% increase in the minimum of $f(x)$ as an error estimate.
}
\label{fig:main-N32-rs3.23-theta1}
\end{figure*}

\begin{figure}
    \centering
    \includegraphics[width=0.9\linewidth]{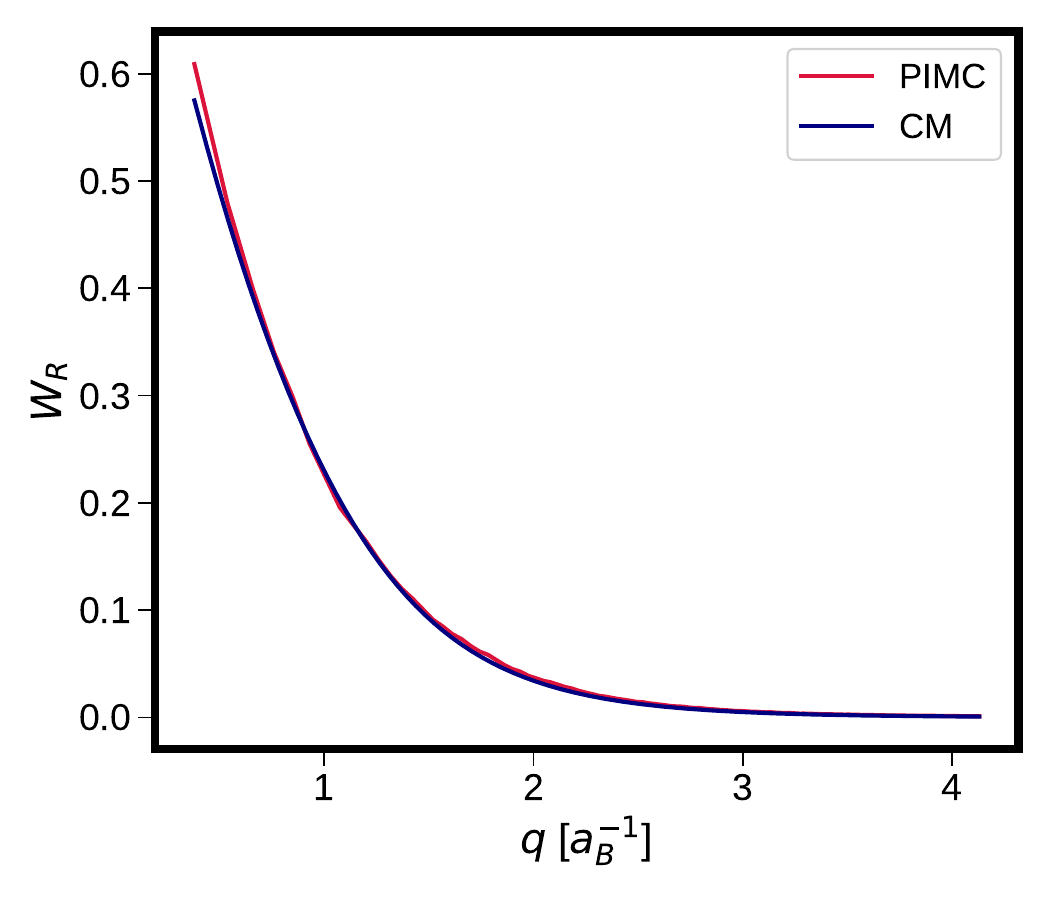}
    \caption{
    Results for the $N=32$, $r_s=3.23$, $\theta=1$ dataset.
    Rayleigh weight $W_R(q)$ compared against PIMC results.}
    \label{fig:main-N32-rs3.23-theta1-Wr}
\end{figure}

In Fig.~\ref{fig:main-N32-rs3.23-theta1} we shot lots of the $\delta$ value for the different fits at various $q$ and $x=Z,\IPD$, for the elastic (top left) and inelastic component (top right).
The corresponding scattering angle $\phi$ for a given scattering number $q$ is plotted on the top for a beam energy of $9\unit{keV}$, which is commonly used in XRTS experiments.
The best fit values over all $q$ are plotted in the dashed red line, whereas the local $\delta$ minima are indicated using the dashed black line.
As shown, we get reasonably consistent values for small $q$, up to $q\sim2\unit{a_B^{-1}}$ for both cases, after which the sensitivity of the $\delta$-error with respect to both IPD and $Z$ drops significantly. 
The bottom panel of Fig.~\ref{fig:main-N32-rs3.23-theta1} shows the sum of $\delta$ over all $q$ values given by
\begin{table}[t]
    \centering
    \setlength{\tabcolsep}{4pt}
    \renewcommand{\arraystretch}{1.1}
        \begin{tabular}{>{\centering\arraybackslash}p{0.06\columnwidth} | >{\centering\arraybackslash}p{0.08\columnwidth} | >{\centering\arraybackslash}p{0.08\columnwidth} | >{\centering\arraybackslash}p{0.13\columnwidth} | >{\centering\arraybackslash}p{0.12\columnwidth} | >{\centering\arraybackslash}p{0.13\columnwidth} | >{\centering\arraybackslash}p{0.15\columnwidth}}
        \textbf{$N$} & \textbf{\emph{$\theta$}} & \textbf{\emph{$r_s$}} & \textbf{$\rho$ }[g/cc] & \textbf{$T_e$ }[eV] & \textbf{$Z^{W_R}$} & \textbf{$\Delta_\text{IPD}^{\text{inel}}[\unit{eV}]$} \\
        \hline
        \hline
        14 & $1$ & $1$     & $2.70$    & $50.11$   & $1.0$     & N/A \\
        14 & $1$ & $2$     & $0.33$    & $12.53$   & $0.66$    & $-1.38$ \\
        14 & $1$ & $2.5$   & $0.17$    & $8.02$    & $0.61$    & $-1.51$ \\
        14 & $1$ & $2.75$  & $0.13$    & $6.63$    & $0.57$    & $-2.47$ \\
        14 & $1$ & $3.0$   & $0.09$    & $5.57$    & $0.51$    & $-3.43$ \\
        14 & $1$ & $3.23$  & $0.08$    & $4.80$    & $0.46$    & $-4.12$ \\
        14 & $2$ & $3.23$  & $0.08$    & $9.61$    & $0.81$    & $-6.32$ \\
        14 & $4$ & $3.23$  & $0.08$    & $19.21$   & $0.96$    & $-13.46$ \\
        14 & $8$ & $3.23$  & $0.08$    & $38.43$   & $0.99$    & $-13.6$ \\
        \hline
        32 & $1$ & $1.0$   & $2.70$    & $50.11$   & $1.0$     & N/A \\
        32 & $1$ & $2.0$   & $0.33$    & $12.53$   & $0.68$    & $-1.37$ \\
        32 & $1$ & $3.23$  & $0.08$    & $4.80$    & $0.45$    & $-4.12$ \\
        32 & $2$ & $3.23$  & $0.08$    & $9.61$    & $0.81$    & $-6.04$ \\
        32 & $4$ & $3.23$  & $0.08$    & $19.21$   & $0.96$    & $-11.81$ \\
        32 & $8$ & $3.23$  & $0.08$    & $38.43$   & $0.99$    & $-13.6$ \\
        \hline
    \end{tabular}
    \caption{
    Fit results over all $q$ for the different datasets available, using an analytic fit to the effective static local field correction derived in \cite{Dornheim_PRB_ESA_2021}.
    For fully ionized cases ($r_s=1$, $\theta=1$) no clear value for the IPD could be found.
}
    \label{tab:fit-results}
\end{table}
\begin{equation}
    f(x) = \sum_i \delta(q_i)
    \,,
\end{equation}
for $x=Z,\IPD$ and normalized arbitrarily to $f(x=0)=1$.
The best fit value for each is indicated by the corresponding dotted line.
We get a clear minimum in $f(Z)$ for the corresponding best fit, indicating a robust value.
This is further emphasized by the good agreement of the fitted elastic component to the PIMC data as shown in Fig.~\ref{fig:main-N32-rs3.23-theta1-Wr}, indicating high confidence in our estimate for the ionization state for the given chemical model.
However, the $f(\IPD)$ plot flattens out around the best fit value of $\IPD=-4.12\unit{eV}$, which indicates that our IPD values are less well defined.
On the other hand, it suggests that the fit is insensitive to the actual value given for the IPD.
This nevertheless demonstrates that in cases where the input parameter has a great impact on the fit, as is apparent for small $q$, we can resolve it well, and in cases where there is little sensitivity, the fit becomes increasingly uncertain.

A summary of the values obtained for all datasets is given in Table~\ref{tab:fit-results}.
Detailed results for each $N=32$ dataset can be found in Appendix~\ref{app:full-results}.
As shown, valid estimates can be obtained over all parameter space considered here, except for the $r_s=1$ and $\theta=1$ case (this is beyond the Mott density \cite{Bonitz_2025}), where the mean interparticle distance is of order the atomic radius, indicating orbital overlap and a regime where the Chihara decomposition and its assumptions of separated free and bound states likely becomes inaccurate.
Further, we observe increasing uncertainty in the IPD estimate for fully ionized cases, where the contribution of the bound-free feature to the full ITCF disappears and the impact of IPD on the fit is negligible.
We also observe consistent results between the $N=14$ and $N=32$ datasets at the same plasma conditions, suggesting finite-size effects are not relevant here~\cite{Chiesa_PRL_2006,dornheim_prl,Dornheim_MRE_2024}; a small exception is given by the IPD extracted from the $r_s=3.23$ and $\theta=4$ datasets.
The likely cause of this is again the insensitivity of the ITCF to the IPD at these parameters, suggesting a relatively high level of uncertainty in the estimated values.
\begin{figure*}
    \centering
    \includegraphics[width=\linewidth]{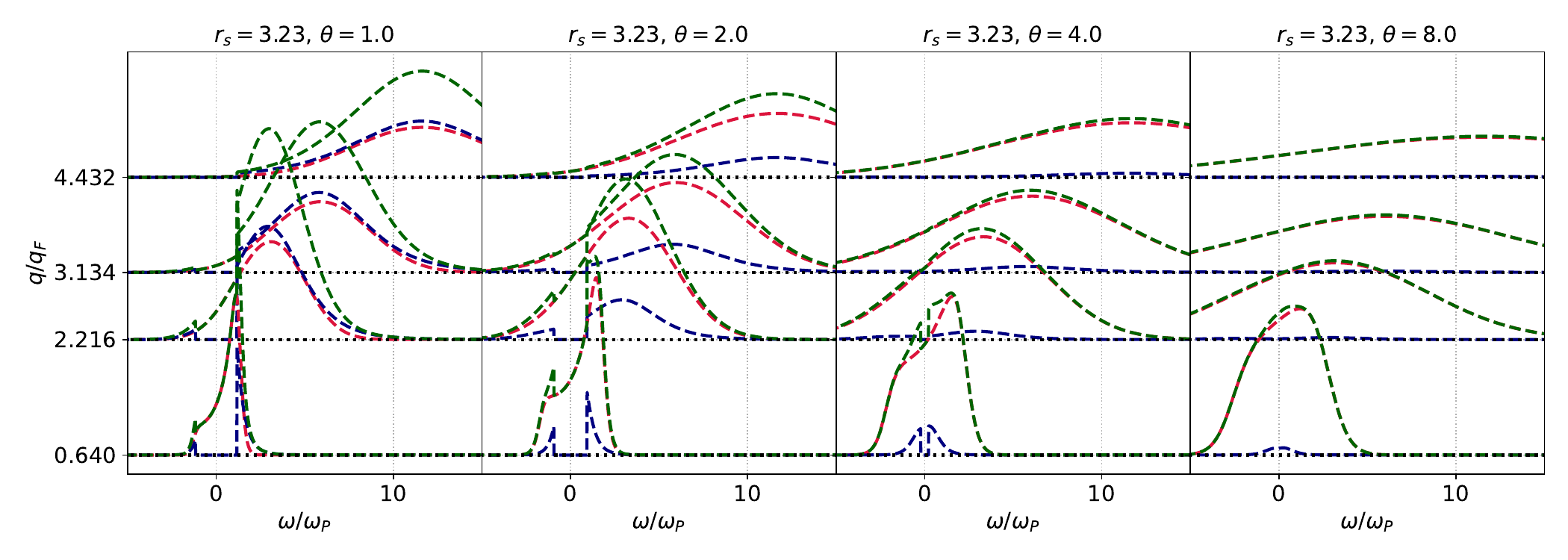}
    \caption{
    Dynamic structure factor for the free-free (red), bound-free (blue) and total (green) interactions at different conditions and increasing $q$ plotted against the energy loss $\omega$ in terms of the plasma frequency $\omega_p$.
    Results are obtained using the best-fit parameters and scaled arbitrarily.
    }
    \label{fig:dsf-tot}
\end{figure*}

We can further investigate the sensitivity of our fit to the ionization state and IPD by having a closer look at the dynamic structure factors obtained using simplified models within the Chihara decomposition.
Our approach not only allows us to directly compare PIMC results against these, but further gives a way of inverting Eq. (~\ref{eqn:laplace-itcf}) to access the dynamic structure factors from the ITCF.
Inverting the Laplace transform is well-known to be exponentially ill-defined~\cite{Epstein_2008}, making the direct extraction of structure factors from the ITCF notoriously difficult.
While recent work has focused on a more accurate algorithm~\cite{chuna2025dualformulationmaximumentropy}, our work here has given us a different approach in directly fitting a DSF to PIMC data using approximate models.
The best fit structure factors are plotted for a selection of datasets in Fig.~\ref{fig:dsf-tot} for an increasing set of $q$ values, with individual components in the Chihara decomposition shown in different colors.
Notably, we see the contribution of the bound-free feature (blue) decreasing at increasing $\theta$, as the system becomes more ionized.
We also observe the width of the DSF increasing for increasing temperature.
Both are expected features in the dynamic structure factor at increasing temperature.
Further, for increasing $q$ the shape of the bound-free DSF increasingly widens, with the effect of the K-edge, and therefore the impact of IPD, disappearing~\cite{Schumacher_1975}.
At $q\sim2.2\unit{a_B^{-1}}$, the form progressively starts resembling the free-free dynamic structure factor.
This would indicate that the ionization simply becomes an arbitrary difference in weighting between the two similar tructure factors for increasing $q$, with its definition gradually becoming less significant. 

\begin{figure*}
    \centering
    \includegraphics[width=0.9\linewidth]{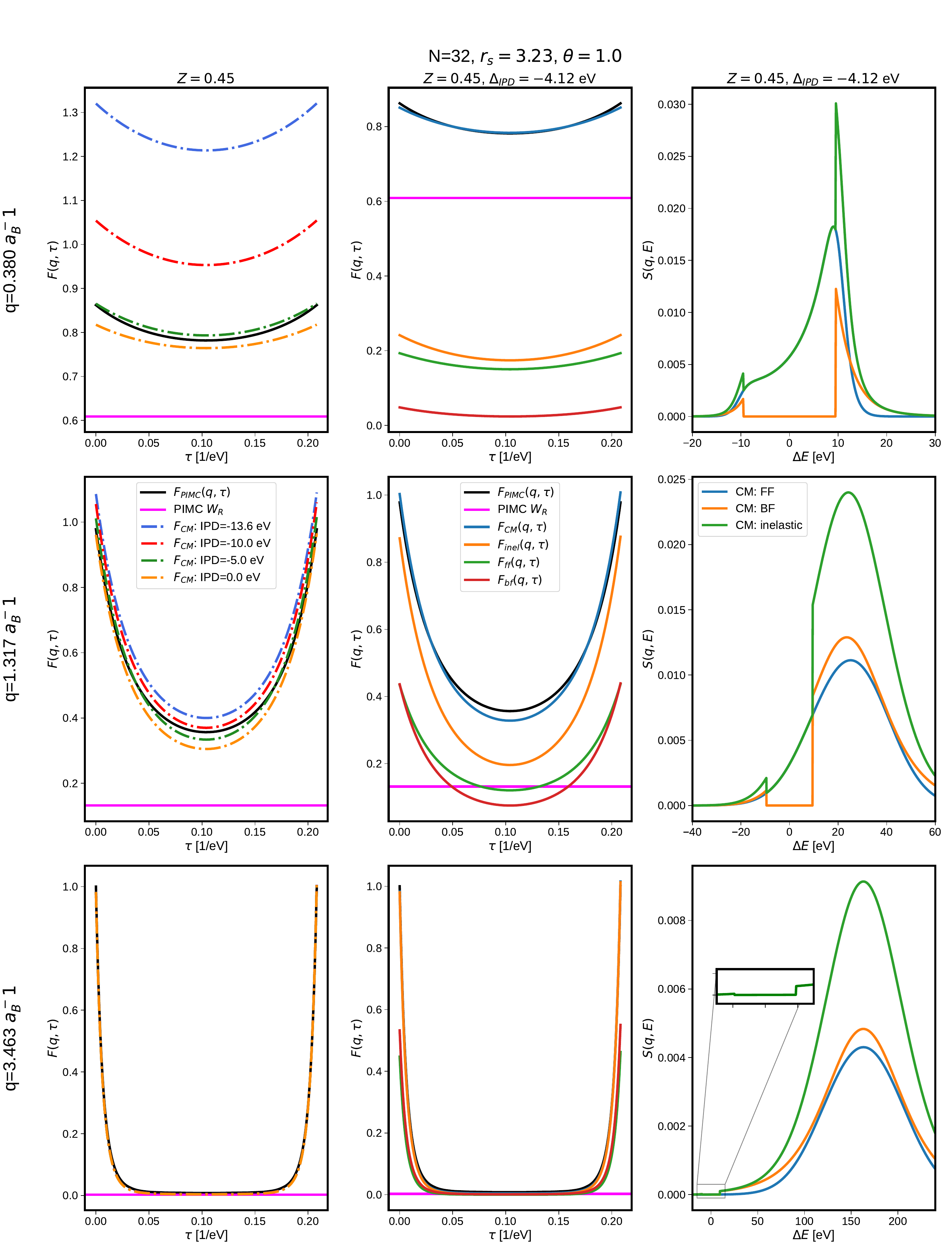}
    \caption{
    Detailed results of the ITCF for $N=32$, $r_s=3.23$ and $\theta=1$ at different $q$, PIMC and chemical model (CM) compared.
    Left column: ITCF for different values of IPD using the best-fit resut for the charge state.
    Middle column: individual components of the ITCF plotted for the best-fit results for both the charge state and the IPD.
    Right column: corresponding structure factors obtained using the chemical model, split up into individual components.
    Top to bottom: increasing $q$, corresponding to increasing scattering angle at constant beam energy. 
    }
    \label{fig:itcf_detailed}
\end{figure*}

A more detailed description of the individual components of the Chihara decomposition in the imaginary time perspective is given by Fig.~\ref{fig:itcf_detailed}.
Three increasing scattering wavenumbers $q$ were selected here for $N=32$, $r_s=3.23$ and $\theta=1.0$, corresponding to hydrogen at $\rho=0.08\unit{g/cm^3}$ and $T_e=4.80\unit{eV}$.
This dataset was chosen because of its best fit ionization state at $Z=0.45$ with a corresponding IPD of $\Delta_\text{IPD}=-4.12\unit{eV}$, showing both the bound-free and the free-free contributions to the inelastic Chihara decomposition in approximately equal relative weighting.
The (constant) Rayleigh weight is plotted in pink for each case as a point of reference.
The left column shows the impact of varying IPD values (dashed lines) on the inelastic ITCF in dotted lines compared against the best fit result.
As demonstrated, the value chosen for the IPD has a direct impact on the shape and position of the ITCF.
An increasing IPD leads to an upshift in the ITCF, thus, as we assume a decreasing IPD (i.e. moving to a more negative value) for more ionized systems, we would expect a similar effect of the ionization state on the ITCF.
This will be investigated in greater detail below.
The middle column shows the individual components of the Chihara decomposition in the ITCF compared against the PIMC result for the best-fit case.
We observe generally good agreement between the PIMC data (black) and the total ITCF (blue), particularly for the low $q$ case seen in the top row.
At medium values of $q$, the fit becomes qualitatively worse.
For high scattering numbers, as can be observed in the bottom panel, the fit starts improving again.
This is likely due to the RPA approximation used here to model the free-free component becoming increasingly accurate in the limit $q \rightarrow \infty$.
The right column shows a plot of the dynamic structure factor obtained using the chemical model in the usual frequency domain for a complete picture.

\begin{figure*}
    \centering
    \includegraphics[width=0.99\linewidth]{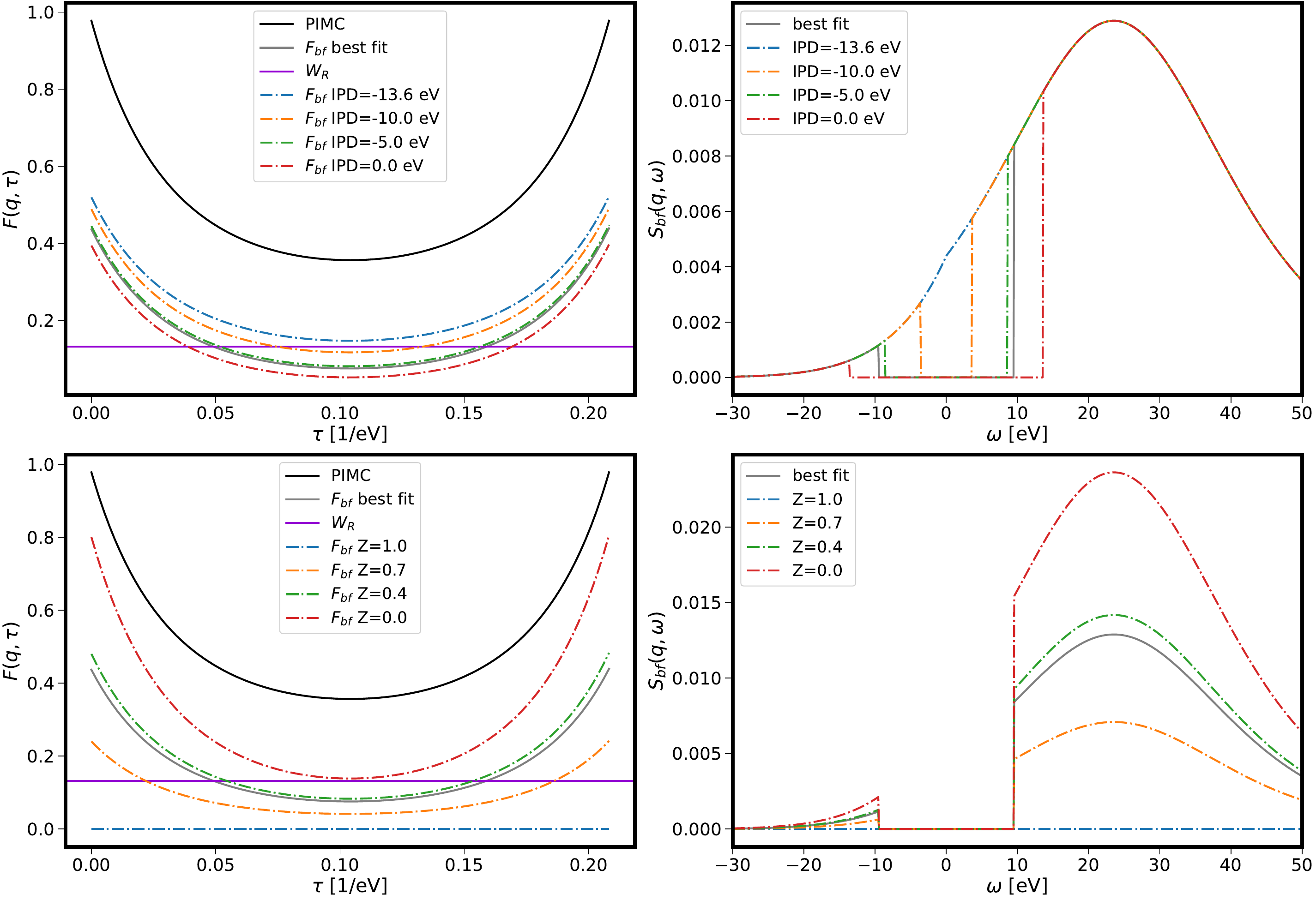}
    \caption{Detailed effect of both ionization and IPD on the bound-free component for a single scattering vector $q=1.3167 \unit{a_B}$ for $N=32$, $r_s=3.23$ and $\theta=1.0$.
    Top row: varying IPD in comparison against the PIMC result.
    Bottom: varying ionization state.
    Left: bound-free component of the ITCF compared against the full PIMC result.
    Right: bound-free dynamic structure factor for different input options.
    }
    \label{fig:bf-detailed}
\end{figure*}

Having considered the impact of varying IPD on the ITCF, let us now extend this analysis to varying ionization states and take a closer look at the bound-free contributions only.
This is shown in Fig.~\ref{fig:bf-detailed}, where we again plotted the bound-free ITCF for varying inputs.
The Rayleigh weight and PIMC result are once again added as a reference point, as are the dynamic structure factors in the frequency domain on the right.
The left column here illustrates the variation in ITCF for different IPD (top) and ionization states (bottom).
The top plot is again a closer look at the left column in Fig.~\ref{fig:itcf_detailed} for a single point in $q$, to demonstrate the impact of IPD on the bound-free component only.
Since the free-free part of the inelastic scattering has no dependence on the IPD, the upwards shift previously observed in the ITCF for decreasing IPD is only due to an up-shift in the bound-free component. 
Because of the nature of the Laplace transform, contributions to the dynamic structure factor at high $\omega$ introduce a steep decay in $\tau$ in the ITCF.
That is to say, any $\tau$-dependency in the ITCF is introduced by shifts or changes in the dynamic structure factor at $\vert \omega \vert > 0$.
Since the IPD only leads to a reduction in the effective binding energy and therefore simply a shift in the K-edges very close to $\omega=0$ for more negative values---as demonstrated in the top right panel showing the bound-free DSF in the usual frequency domain---we only observe an effectively constant shift in the ITCF due to the increased contributions of the DSF.
Constructing a similar analysis for the ionization state, we can further investigate the influence of varying $Z$ for a fixed IPD on the bound-free ITCF.
As illustrated in the bound-free structure factor in the usual frequency domain (right), increasing $Z$ simply from $Z=0$ for the fully bound to $Z=1$ for the fully ionized case simply introduces a multiplying factor of $1/(1-Z)$ to the feature.
This is a result of the impulse approximation used here being formulated as a sum over the contributions of each sub-shell~\cite{Schumacher_1975} and scaled using the relative number of bound electrons defined as $Z_b = 1 - Z$.
The scaling factor due to the ionization state manifests itself similarly in the ITCF, although we see a clear $\tau$-dependence that is introduced by the $e^{-\tau\omega}$ factor in the Laplace transform.
Notably, we observe different impacts of ionization and IPD on the shape of the ITCF, but it further demonstrates that simultaneously optimizing for both IPD and ionization here would introduce considerable co-dependencies, and illustrates the need for carefully considering the relationship between both.

\begin{figure}
    \centering
    \includegraphics[width=0.95\linewidth]{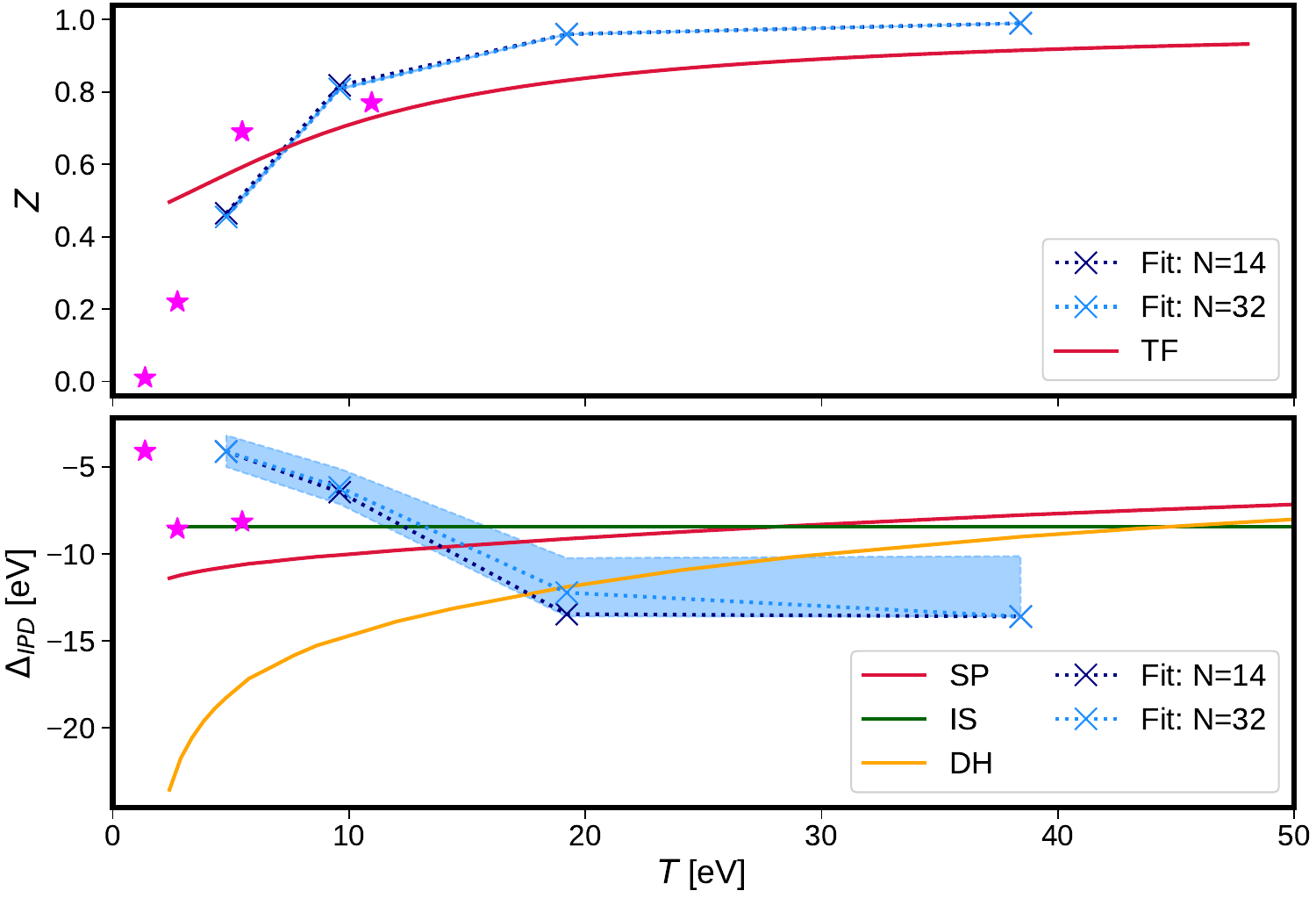}
    \caption{   
    Comparison of best-fit values for a fixed $r_s=3.23$ corresponding to $\rho=0.08\unit{g/cc}$.
    Top: best-fit ionisation state plotted against temperature in comparison with the commonly used Thomas-Fermi (TF) model.
    The magenta datapoints indicate fermionic PIMC ionization results obtained by Filinov and Bonitz~\cite{filinov2023equation}.
    Bottom: IPD compared against the Stewart-Pyatt (SP), Debye-H\"uckel (DH) and Ion-sphere (IS) models using the estimated TF ionization degree.
    Corresponding results using our estimated ionization degree are given in~\ref{fig:fit-comparison-new}.
    Fit results from the $N=14$ and $N=32$ datasets are plotted separately in light and dark blue points.
    The magenta points are a comparison to the IPD by Bonitz and Kordts~\cite{Bonitz_2025}.
    Dotted lines are drawn as a guide to the eye.
    The shaded areas indicate the error estimated using a 10\% deviation from the minimum found for the $N=32$ dataset as indicated by the horizontal lines in Fig.~\ref{fig:main-N32-rs3.23-theta1}.
}
    \label{fig:fit-comparison}
\end{figure}

\begin{figure}
    \centering
    \includegraphics[width=0.95\linewidth]{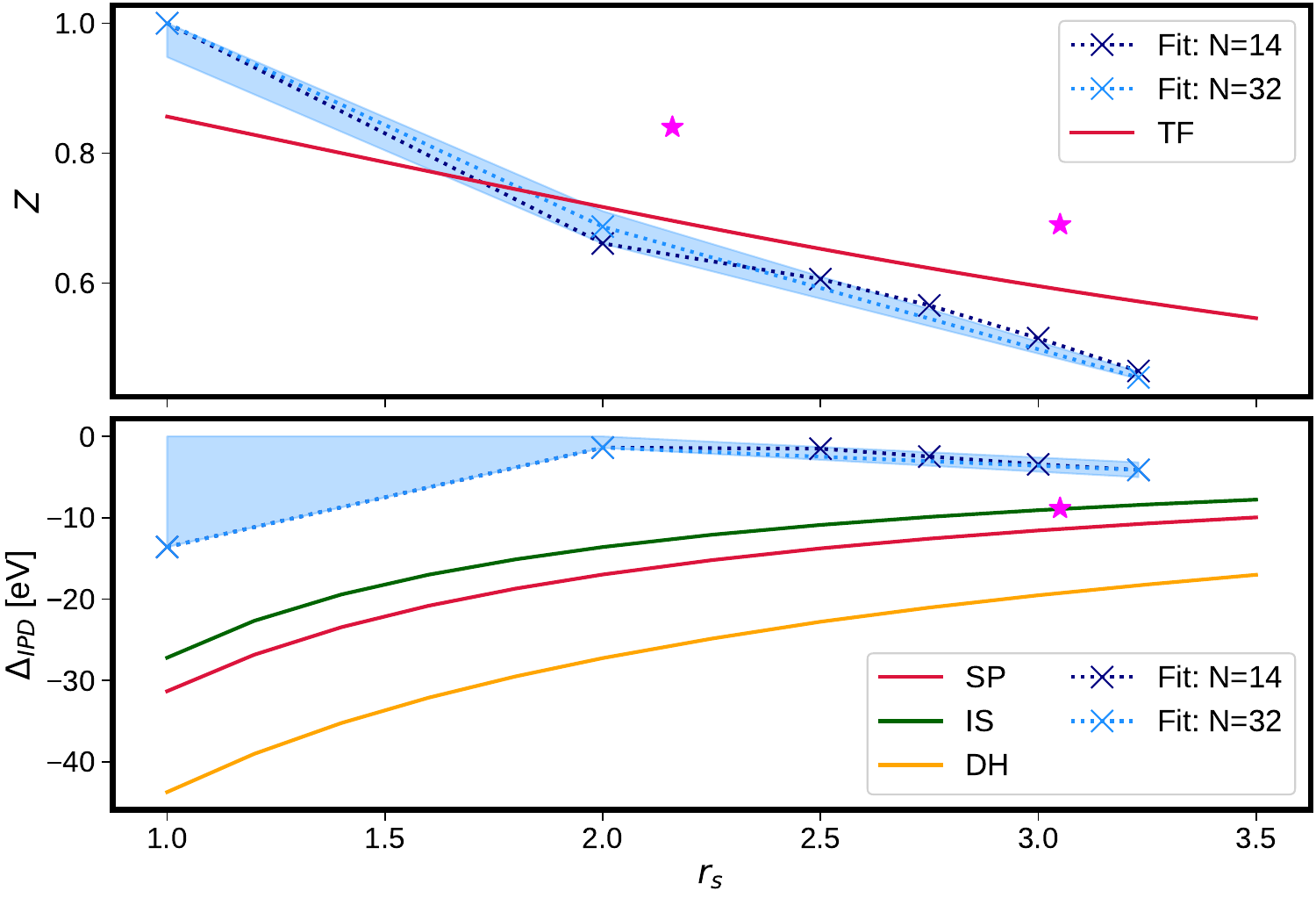}
    \caption{Comparison of best-fit values for a fixed $\theta=1$.
    Top: best-fit ionization state plotted against temperature in comparison with the commonly used Thomas-Fermi (TF) model. 
    Bottom: IPD compared against the Stewart-Pyatt (SP), Debye-H\"uckel (DH) and Ion-sphere (IS) models using the estimated TF ionization degree.
    Corresponding results using our estimated ionization degree are given in~\ref{fig:fit-comparison-const-theta-new}.
    Fit results from the $N=14$ and $N=32$ datasets are plotted separately in light and dark blue points.
    Dotted lines are drawn as a guide to the eye.
    The shaded areas indicate the error estimated using a 10\% deviation from the minimum found for the $N=32$ dataset as indicated by the horizontal lines in Fig.~\ref{fig:main-N32-rs3.23-theta1}.
    More points of comparison that confirm the general trend for higher $r_s$ can be found in the work of Bonitz and Kordts~\cite{Bonitz_2025}.
}
    \label{fig:fit-comparison-const-theta}
\end{figure}

Figures~\ref{fig:fit-comparison} and~\ref{fig:fit-comparison-const-theta} show the comparison of models most commonly used in the construction of equation of state tables against subsets of the fitted results.
The Thomas-Fermi ionization model here is taken from \cite{Murillo_2013,Stanton_2018} for a multi-component system.
Commonly used models for the IPD, such as Stewart-Pyatt (SP), Ion-sphere (IS) and Debye-H\"uckel are plotted for comparison.
The Debye-H\"uckel model~\cite{Hückel_1924} presents the simplest IPD model here, and is only valid in the weakly coupled limit, whereas the ion-sphere model assumes a strong coupling between the ions \cite{Crowley_2014_ipd}.
One of the most commonly used IPD models, the Stewart-Pyatt model, acts as an interpolation between both limits~\cite{Stewart_1966}.
Two cases are considered here, each covering a subset of the available PIMC datasets: for constant $r_s=3.23$ (see Fig.~\ref{fig:fit-comparison}) and constant $\theta=1.0$ (see Fig.~\ref{fig:fit-comparison-const-theta}).
The constant $r_s$ case allows a direct comparison of the IPD at increasing temperature, whereas the constant $\theta$ case lies closely to the path traversed by a typical ICF capsule during implosion in $\rho-T$ space \cite{hu_ICF,Dornheim_review} and thus allows us to directly study the impact of this work on relevant applications.
In both cases, the best-fit ionization derived here, generally follows the trend of the Thomas-Fermi model, but predicts higher values for the ionization state at high temperatures and lower values at low densities.
For completeness, we have also added two reference points from Filinov and Bonitz~\cite{filinov2023equation} for the ionization degree that have been obtained from PIMC simulations via a cluster analysis.
Subsequently, Bonitz and Kordts~\cite{Bonitz_2025} have extracted a corresponding IPD which has been included in the bottom panel of Fig.~\ref{fig:fit-comparison}.
These results are of the same order of magnitude compared to the present work, although some quantitative differences appear.
The comparison against commonly used IPD models is similar, where the constant $\theta$ case shows the same general trend and agreement to the IS and DH models at low density.
For the constant $r_s$ case however, we see some divergent behavior at increasing temperatures.
While all three models predict an increasing IPD at $T>20\unit{eV}$, our results predict a continuous decrease until the fully ionized case.
This would indicate that our approach works well in regimes where the ITCF is very sensitive to either the IPD or the ionization state, but becomes increasingly uncertain in regimes where the sensitivity drops.
We observe that this is likely due to similar behavior of the optimization function as observed in Fig.~\ref{fig:main-N32-rs3.23-theta1}, which is very flat around the proposed $\delta$ minimum, indicating that the value has very little impact.
The shaded areas indicate a 10\% increase in the $f(x)$ around the minimum, as shown in the bottom panel of~\ref{fig:main-N32-rs3.23-theta1} by the dotted gray lines, to obtain an error estimate, further showing the increasing uncertainty of our fit due to the decreasing sensitivity for highly ionized systems.
This is expected, since for the almost fully ionized case the contribution of the bound-free component to the total ITCF is very small, and thus any change in the value for the IPD would have little effect on the result.
We have plotted a similar comparison using the estimated ionization degree as input to the IPD model in Figures~\ref{fig:fit-comparison-new} and~\ref{fig:fit-comparison-const-theta-new} in Appendix~\ref{app:full-results}.

Several studies in the past have evaluated the validity of each of these models (e.g.~\cite{Crowley_2014_ipd,Ciricosta_2012,Gawne_2023}).
Experimental measurements of the continuum lowering effect in Aluminum have been found to either match Stewart-Pyatt or deviate from it significantly~\cite{Crowley_2014_ipd}, clearly indicating that further development is needed to arrive at an accurate continuum lowering description.
The Crowley IPD model~\cite{Crowley_2014_ipd} would offer another point of comparison that has been shown to match experimental data better.

The most notable observation however comes from Fig.~\ref{fig:main-N32-rs3.23-theta1} and the impacts of IPD on the ITCF.
For both the IPD and the mean ionization state, we observe a steep drop-off in sensitivity at increasing $q$.
As seen in Fig.~\ref{fig:main-N32-rs3.23-theta1}, any sensitivity of $\delta$ with respect to the IPD for the inelastic component disappears at about $q\approx2.5 \unit{a_B}$.
We observe similar behavior when considering the sensitivity of the inelastic feature with respect to the ionization state.
Further, any sensitivity of the elastic feature with respect to the ionization state largely disappears at $q\approx3.0\unit{a_B}$.
This would be somewhat expected, since the Rayleigh weight is exponentially reducing at increasing $q$, see Fig.~\ref{fig:main-N32-rs3.23-theta1-Wr}, as the ionic contributions to the scattering disappear in the non-collective regime.
For a beam energy of $9\unit{keV}$, commonly used in XRTS experiments, this corresponds to a scattering angle of $\theta_{\text{sc}} = 117 ^\circ$.
This would indicate, that we can indeed find consistent results over all scattering numbers for both the IPD and the ionization, given that the sensitivity of both decreases with increasing $q$.
We have demonstrated this by finding a single best fit value for both input parameters for all $q$ (see Fig.~\ref{fig:main-N32-rs3.23-theta1}), which agrees well with the smaller scattering numbers and only disagrees for increasing $q$ where the actual value is shown to have little impact on the shape of both the ITCF and the dynamic structure factors.

Intuitively, this can be explained by considering concepts like ionization and IPD in the single-particle scattering regime.
The scattering parameter can be defined in terms of $q$ as \cite{siegfried_review}:
\begin{equation}
    \alpha = \frac{1}{q \lambda_s}
    \,,
\end{equation}
using the Debye length to define the plasma screening length $\lambda_s=\kappa_\text{De}^{-1}$.
The parameter $\alpha$ is commonly used to distinguish between the non-collective and collective regimes, with the single-particle regime characterized by $\alpha \ll1 $ and the collective regime being defined by $\alpha \gg1$.
Here, for very dense and very hot systems, we observe a slightly later drop-off in the sensitivity.
For the lowest density case of $r_s=3.23$, the transition between non-collective and collective regime occurs at a smaller scattering wavenumber compared to the highest density case of $r_s = 1$.
For the $r_s=3.23$, $\theta=1.0$ case, it occurs at around $q \sim 0.33\unit{a_B}$.
Conversely, for the higher density, higher temperature case ($r_s=1.0$, $\theta=1.0$), the transition point is at $q \sim 1.27\unit{a_B}$.
The non-collective regime, by definition, probes single particle photon-electron exchanges, therefore probing on much smaller length scales, where averaging effects like the ionization - here defined as the mean number of free electrons per hydrogen nucleus - or screening effects like IPD increasingly have no effect. 

We would thus counsel against drawing a definitive kind of conclusion from the inelastic component of the backward scattering regime in the interpretation of XRTS experiments regarding either IPD or ionization state in the lower density/ temperature regime.
While backwards scattering spectra can for example indicate the position of a K-edge, this becomes increasingly meaningless for larger scattering angles, as the bound electrons observed in the single-particle limit start behaving more like free electrons~\cite{Schumacher_1975}.
Indeed, we can observe this behavior for large $q$ in the top panel of Fig.~\ref{fig:dsf-tot}, where the free-free and bound-free components of the dynamic structure factor start resembling one another in shape.
For increasing $q$, we see any effect of the continuum lowering on the bound-free interactions disappear entirely.

\section{Summary and Outlook\label{sec:summary}} 

In this work, we have demonstrated an approach to diagnose ionization state and continuum lowering from \emph{ab initio} path integral Monte Carlo simulations of warm dense hydrogen by analyzing the imaginary-time correlation function using the Chihara decomposition.
By applying this chemical model to exact quantum simulations, we are able to clearly define concepts like ionization state and continuum lowering via the mean charge state and ionization potential depression respectively. Both of these definitions arise from a fundamental assumption in the Chihara decomposition that electrons are well-distinguished as being either bound to an ion or completely free (i.e. the chemical picture).
It should be noted that this is a simplified description of real electronic states in  high energy density systems, which in general can also be ``quasi-bound`` and difficult to distinguish~\cite{Gawne_2023}.
However, the conceptual simplicity and computational efficiency of the chemical picture, particularly in describing ill-defined concepts such as ionization, makes it nevertheless a useful approximation for studying high energy density systems. 
Using a forward-fitting approach commonly employed to analyze XRTS spectra, we fitted a chemical model to the exact data instead of experimental measurements, allowing us to obtain measurements without any experimental uncertainty or the difficulty that comes with hydrogen's low scattering cross section.
We have shown that we can obtain consistent results for both IPD and ionization state over a large parameter space in both density and temperature, and over a wide range of scattering scattering angles, with the fitting procedure only failing for highly degenerate cases with overlapping orbitals.
A comparison against commonly used ionization and IPD models is given as a reference point, though the validity of these models in the warm dense matter regime needs to be explored further.
We show that results produced here follow the same general trend for both ionization and IPD, with deviations at very high temperature and low density.
A more extensive review of continuum lowering \cite{Crowley_2014_ipd} and the comparison against experimental measurements of the IPD for different materials (see e.g. \cite{Vinko_2015}) have in the past shown however, that these simplified models often fail to accurately describe the complex interactions of the warm dense matter regime, clearly indicating the need for different approaches using exact data.
What our approach here has shown is that we can extract estimates from exact data with good result for parameters where the ITCF is sensitive to variations in either $Z$ or IPD, and only fails when the value has been show to have an indistinguishable effect on the fit.

The mathematical inversion of the Laplace transform is a notoriously difficult problem~\cite{Epstein_2008}, thus directly extracting structure factors from PIMC data has thus far been a challenge.
More recent work has focused on extracting structure factors from uniform electron gas PIMC data~\cite{chuna2025dualformulationmaximumentropy}, paving the first steps in directly extracting a DSF from exact quantum Monte Carlo simulations.
Here, by comparing PIMC simulations to simplified models of the Chihara decomposition, describing bound-free and free-free components separately, we effectively invert Eq. (~\ref{eqn:laplace-itcf}) and get qualitative insights from exact simulations into the dynamic structure factors.
We can also directly investigate the impact of variables commonly measured in XRTS experiments, like ionization and IPD.

By investigating the Chihara decomposition in the imaginary-time domain, we can demonstrate the direct impact both IPD and $Z$ have on the inelastic component of the dynamic structure factor.
Further, this approach has allowed us to test the Chihara decomposition against exact data over a large range of scattering vectors $q$ and has allowed us to demonstrate the abrupt drop in sensitivity of the inelastic component of the dynamic structure factor with respect to both ionization and IPD.
Therefore, one can deduce that the extraction of an ionization state or even IPD from a given XRTS spectrum becomes increasingly meaningless with increasing scattering angle.
While we show that this behavior is density and temperature dependent, we caution against extracting an ionization state from a non-collective XRTS result for future experiments and advise careful consideration of the length scales probed.

Let us briefly discuss the degree of model-dependency regarding the extracted ionization degree and IPD.
First, for the computation of the different components in the Chihara formula we have used common approximations such as the Debye-H\"uckel pseudo-potential and the impulse approximation for the bound-free feature.
In addition, we note that we have not distinguished between different atomic bound states.
This is well justified for the densities and temperatures considered in the present work ($T\gtrsim 3 \unit{eV}$ or $r_s \lesssim 2$), but might be investigated in dedicated future works.

While this approach has so far only been applied to hydrogen, it can be extended to any material or indeed any \emph{ab initio} simulation tool like DFT.
We intend to apply a similar approach to warm dense Beryllium~\cite{Dornheim_Science_2024,Tilo_Nature_2023,Dornheim_2024_preprint} to continue studying the use of the Chihara decomposition as a tool for analyzing PIMC data and further explore its limitations.
Further, the $\rho-T$ parameter space investigated here corresponds to parameters currently accessible in hydrogen jet XRTS experiments at x-ray free electron laser facilities like the European XFEL and LCLS~\cite{Fletcher_2016}, allowing us to validate results extracted here against experimental data.
The advantage of our method here is that it does not rely on any further models like the Saha equation or a Thomas-Fermi ionization model to derive either IPD or ionization state.
What is commonly done in the construction of EOS models, where a relationship between IPD and ionization is defined to derive one from the other, is not necessary here. 
Instead, we have shown that we can derive consistent values for both from the same dataset without introducing any co-dependencies.
Indeed, with the continued advances in \emph{ab initio} modeling and extensions of PIMC capabilities to model increasingly heavier elements, we believe that this is only the start of increasingly accurate descriptions of both IPD and $Z$ by making use of exact data.
We have observed some systematic deviations in the ITCF between the best-fit chemical models and PIMC, particularly for moderate values of the scattering number, indicating that further work is required in optimizing these models to the warm dense matter regime.
Recently published work on the analytic continuation~\cite{chuna2025dualformulationmaximumentropy,dornheim_dynamic,JARRELL1996133} will further allow direct access to structure factors from PIMC, paving the way towards improving models that can be used for the analysis of experiments and construction of EOS tables using exact data.



\begin{acknowledgments}
This work was partially supported by the Center for Advanced Systems Understanding (CASUS), financed by Germany’s Federal Ministry of Education and Research (BMBF) and the Saxon state government out of the State budget approved by the Saxon State Parliament.
This work has received funding from the European Union's Just Transition Fund (JTF) within the project \emph{R\"ontgenlaser-Optimierung der Laserfusion} (ROLF), contract number 5086999001, co-financed by the Saxon state government out of the State budget approved by the Saxon State Parliament.
This work has received funding from the European Research Council (ERC) under the European Union’s Horizon 2022 research and innovation programme
(Grant agreement No. 101076233, "PREXTREME"). 
Views and opinions expressed are however those of the authors only and do not necessarily reflect those of the European Union or the European Research Council Executive Agency. Neither the European Union nor the granting authority can be held responsible for them. Computations were performed on a Bull Cluster at the Center for Information Services and High-Performance Computing (ZIH) at Technische Universit\"at Dresden, at the Norddeutscher Verbund f\"ur Hoch- und H\"ochstleistungsrechnen (HLRN) under grant mvp00024, and on the HoreKa supercomputer funded by the Ministry of Science, Research and the Arts Baden-W\"urttemberg and by the Federal Ministry of Education and Research.
The work of M.~P.~B.~ and Ti.~D. was performed under the auspices of the U.S. Department of Energy by Lawrence Livermore National Laboratory under Contract No. DE-AC52-07NA27344 and supported by Laboratory Directed Research and Development (LDRD) Grant Nos. 24-ERD-044 and 25-ERD-047.
\end{acknowledgments}


\bibliography{bibliography}
\clearpage

\appendix
\section{Full results}\label{app:full-results}

All simulations were run using the analytic approximation of the effective static local field correction published by Dornheim et al.~\cite{Dornheim_PRB_ESA_2021,Dornheim_PRB_2021}, using the Impulse approximation~\cite{Schumacher_1975} to model the bound-free component, a numerical RPA for the free-free, Debye-Huckel for the ion-ion interactions, finite wavelength model for the screening cloud and effective Coulomb to describe electron-ion interactions~\cite{siegfried_review}.
The detailed results for each $N=32$ dataset are given in the plots below.

\begin{figure}
    \centering
    \includegraphics[width=0.95\linewidth]{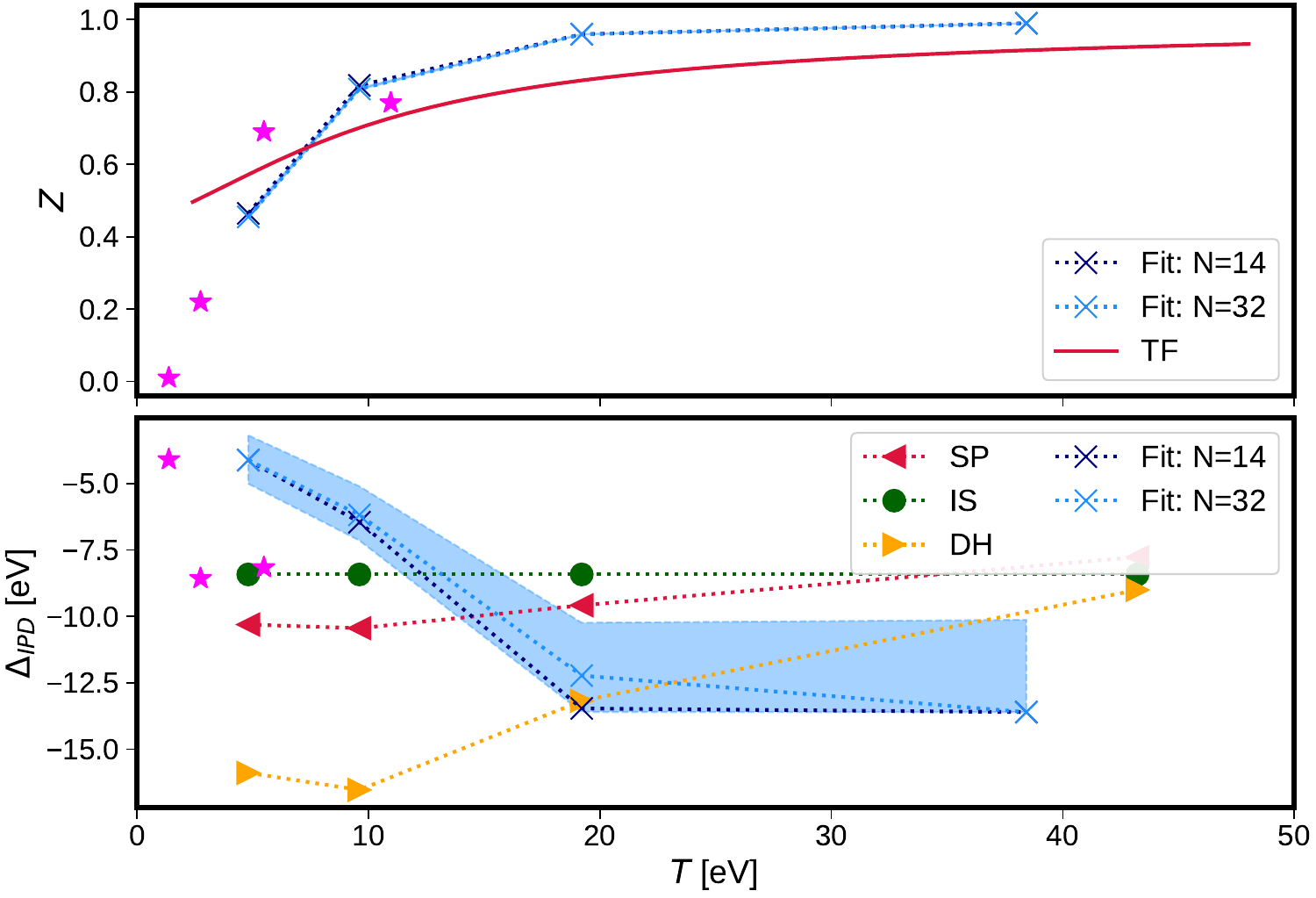}
    \caption{
    Comparison of best-fit values for a fixed $r_s=3.23$ corresponding to $\rho=0.08\unit{g/cc}$.
    Top: best-fit ionisation state plotted against temperature in comparison with the commonly used Thomas-Fermi (TF) model.
    The magenta datapoints indicate ionization results obtained by Filinov and Bonitz~\cite{filinov2023equation}.
    Bottom: IPD compared against the Stewart-Pyatt (SP), Debye-H\"uckel (DH) and Ion-sphere (IS) models using the fitted ionization degrees.
    Fit results from the $N=14$ and $N=32$ datasets are plotted separately in light and dark blue points.
    The magenta point is a comparison to the IPD Bonitz and Kordts~\cite{Bonitz_2025}.
    Dotted lines are drawn as a guide to the eye.
    The shaded areas indicate the error estimated using a 10\% deviation from the minimum found for the $N=32$ dataset.
 }
    \label{fig:fit-comparison-new}
\end{figure}

\begin{figure}
    \centering
    \includegraphics[width=0.95\linewidth]{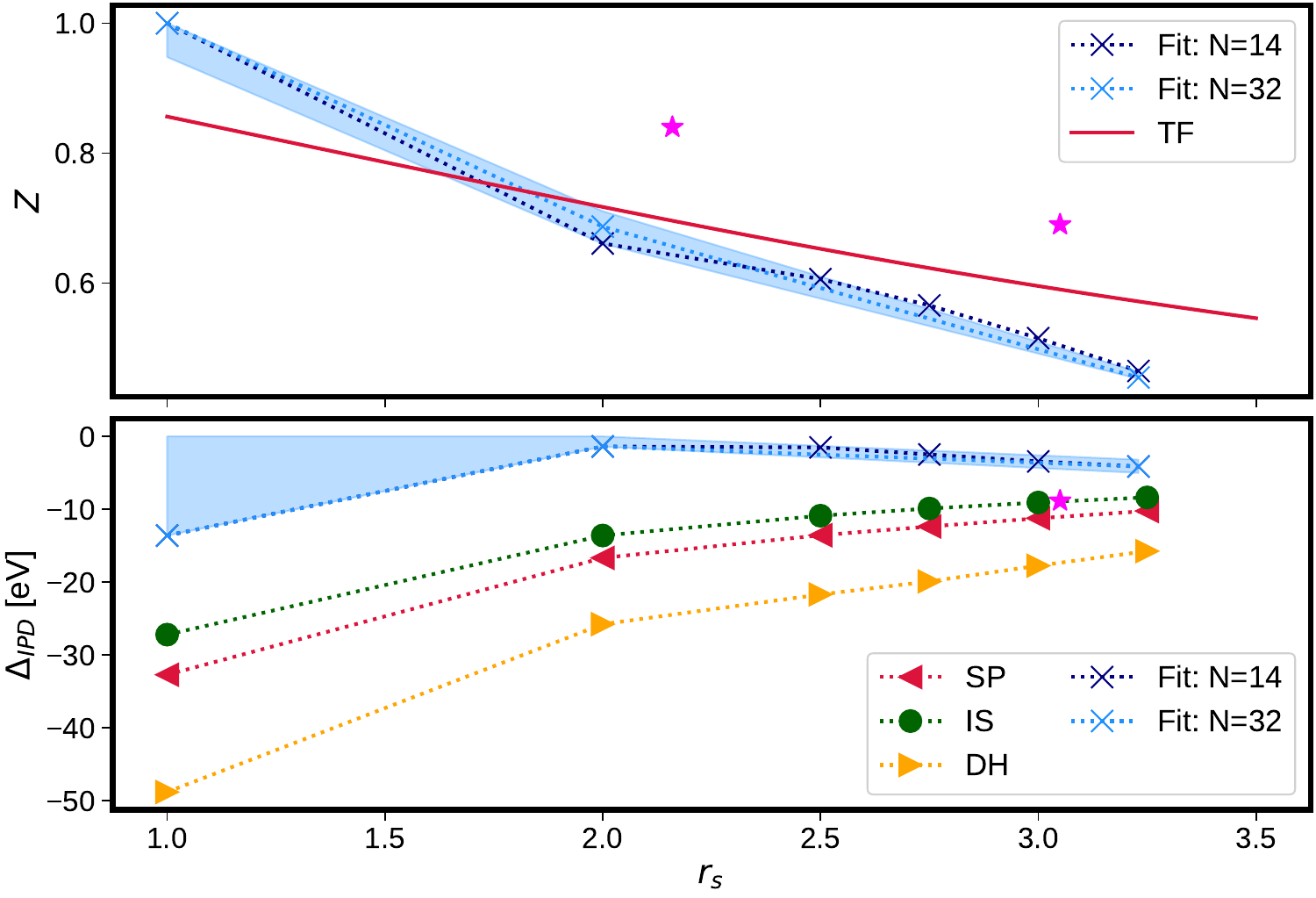}
    \caption{Comparison of best-fit values for a fixed $\theta=1$.
    Top: best-fit ionization state plotted against temperature in comparison with the commonly used Thomas-Fermi (TF) model. 
    Bottom: IPD compared against the Stewart-Pyatt (SP), Debye-H\"uckel (DH) and Ion-sphere (IS) models using the fitted ionization degrees.
    Fit results from the $N=14$ and $N=32$ datasets are plotted separately in light and dark blue points.
    Dotted lines are drawn as a guide to the eye.
    The shaded areas indicate the error estimated using a 10\% deviation from the minimum found for the $N=32$ dataset.
    }
    \label{fig:fit-comparison-const-theta-new}
\end{figure}

Fig.~\ref{fig:fit-comparison-new} and Fig.~\ref{fig:fit-comparison-const-theta-new} also give a comparison of the obtained IPD estimates against commonly used models for the input ionization degree obtained here, as opposed to the Thomas-Fermi model.

\begin{figure*}[ht]
    \includegraphics[width=0.48\linewidth]{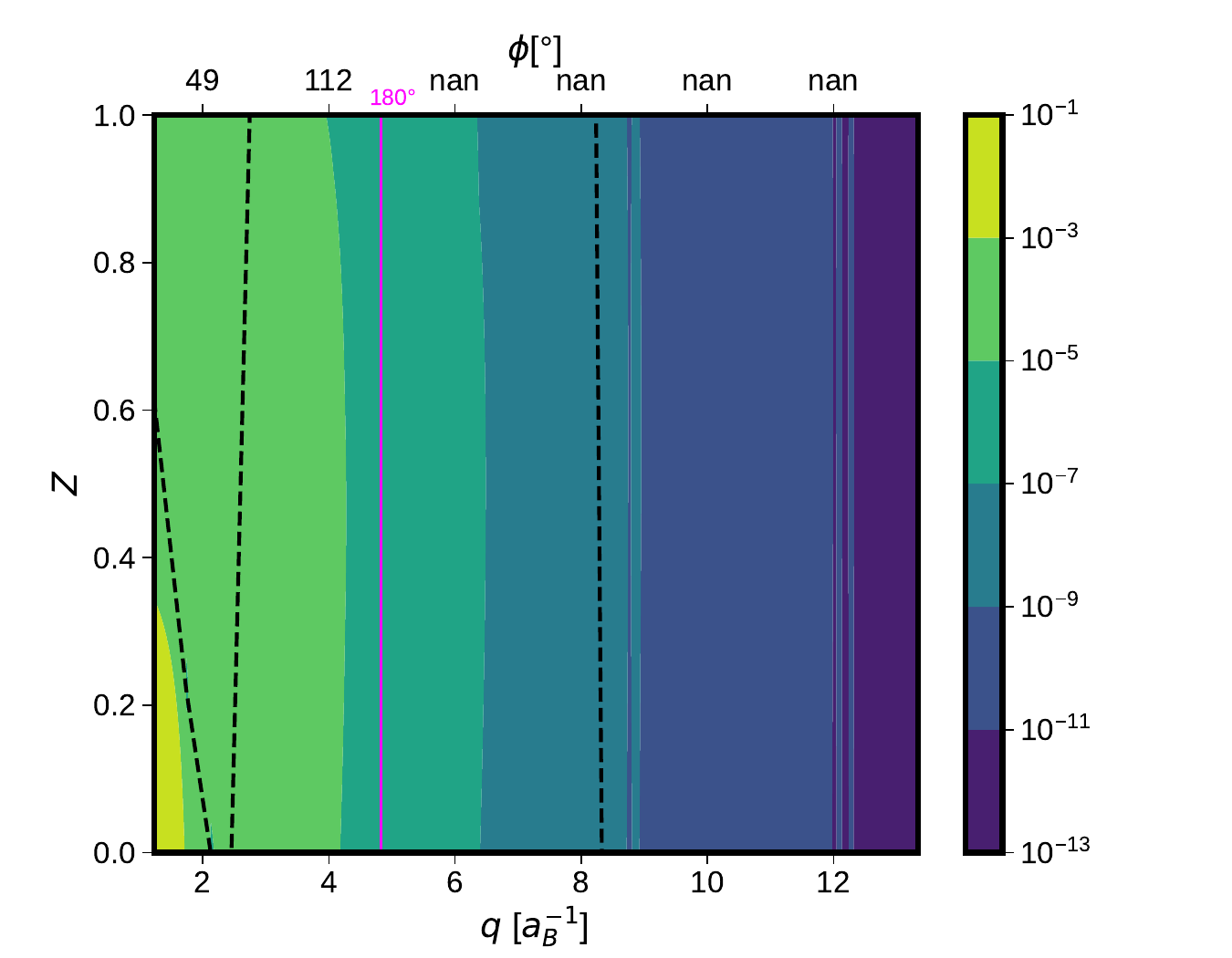}
    \includegraphics[width=0.48\linewidth]{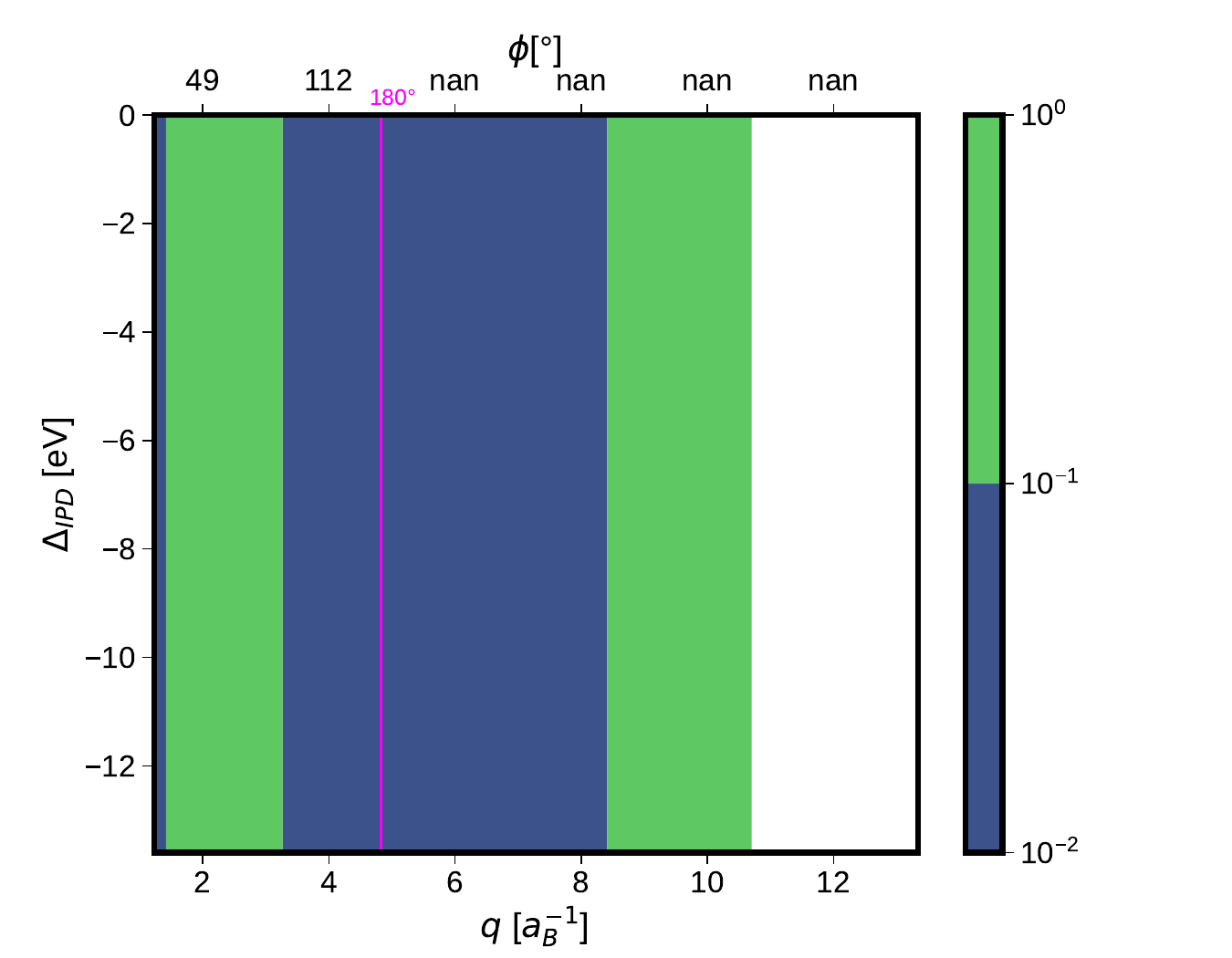}
    \includegraphics[width=0.48\linewidth]{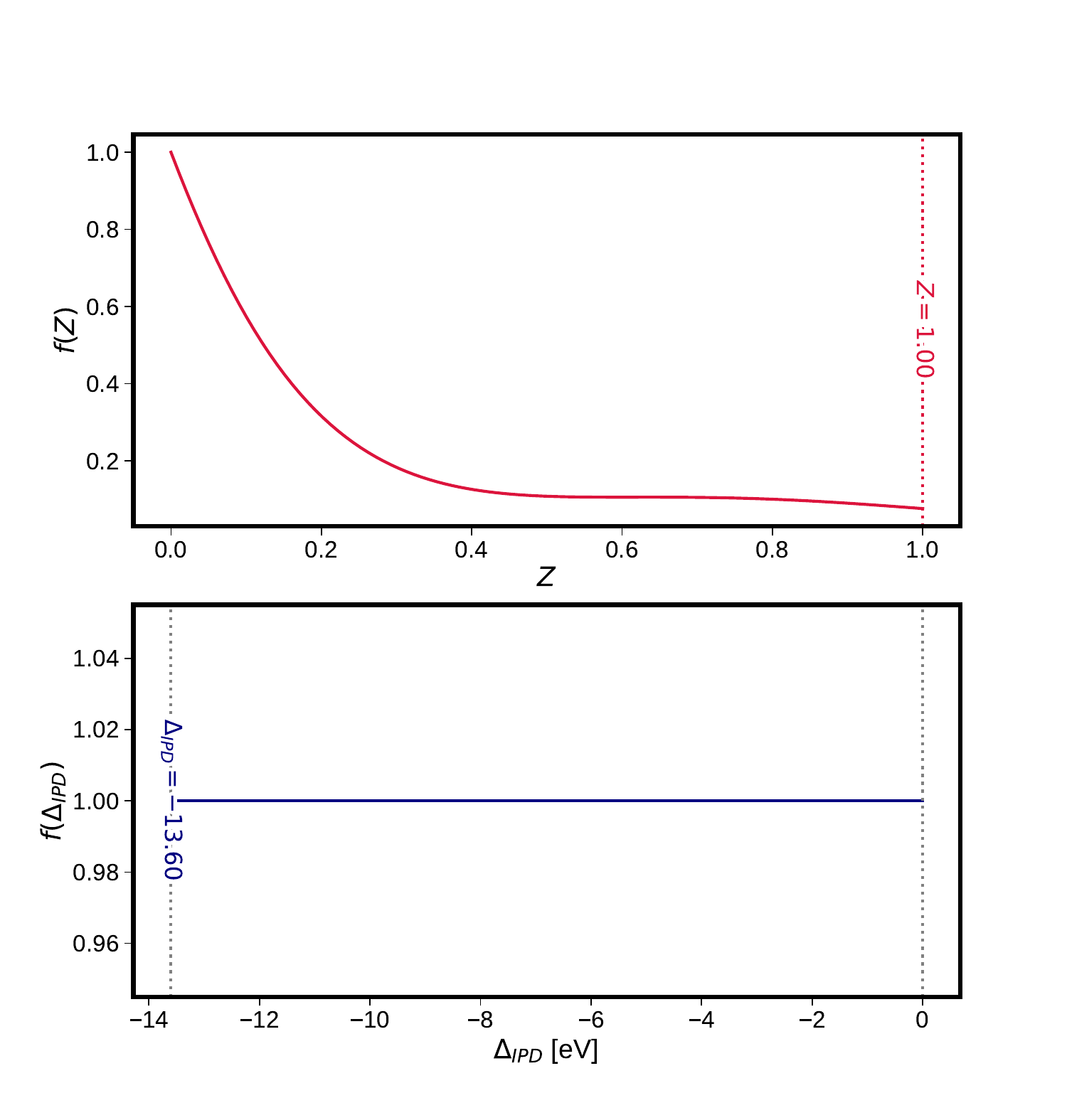}
    \includegraphics[width=0.46\linewidth]{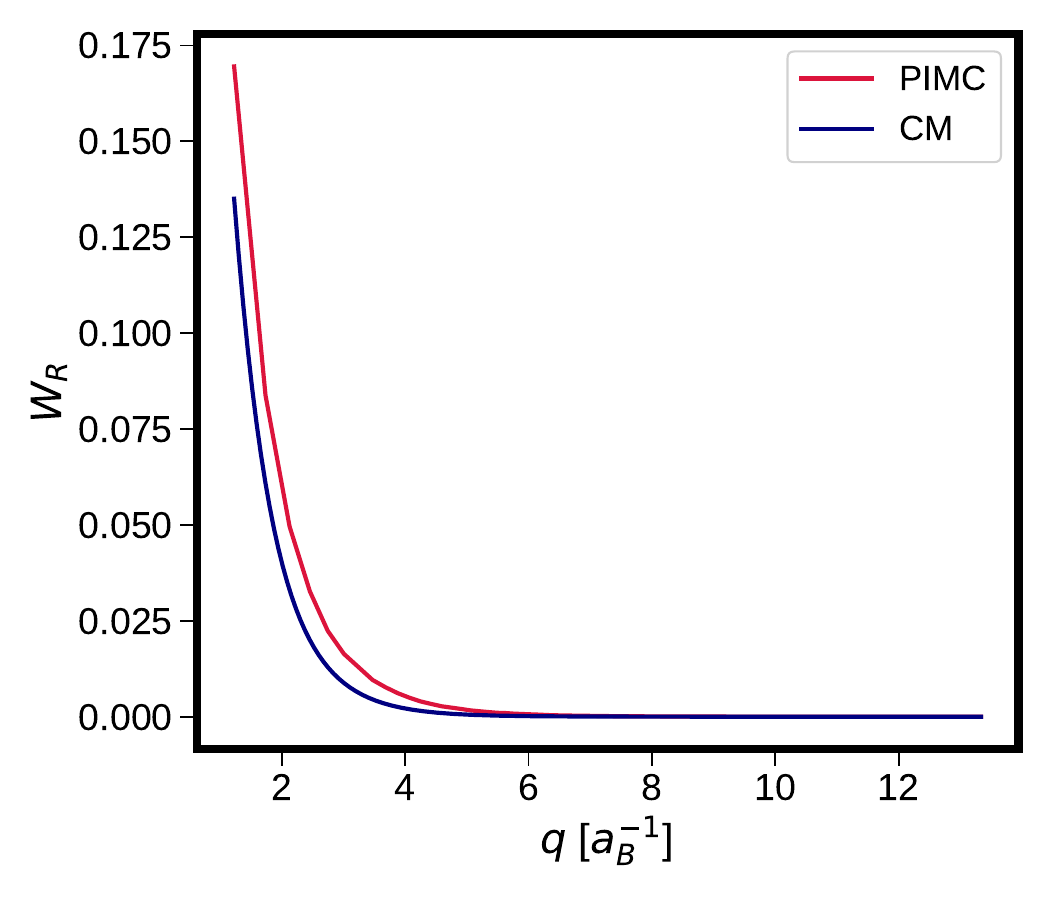}
    \caption{Results $N=32$, $r_s=1$, $\theta=1$ dataset. 
    Top left: $\chi$-error plot for the Rayleigh weight over $q-Z$ parameter space.
    Top right: $\chi$-error plot for the inelastic component over $q-\Delta_{\text{IPD}}$ parameter space.
    For each surface plot, corresponding scattering angles $\phi$ for a beam energy of $9\unit{keV}$ are shown on the top x-axis.
    The $180^{\circ}$ cut-off is shown in the solid pink horizontal line.
    Bottom left: ionization and IPD error, arbitrarily normalized to enforce $f(Z=0)=1$.
    Bottom right: best fit $W_R(q)$ plotted against the PIMC results.
    }
    \label{fig:N32-rs1-theta1}
\end{figure*}

\begin{figure*}[h]
    \includegraphics[width=0.48\linewidth]{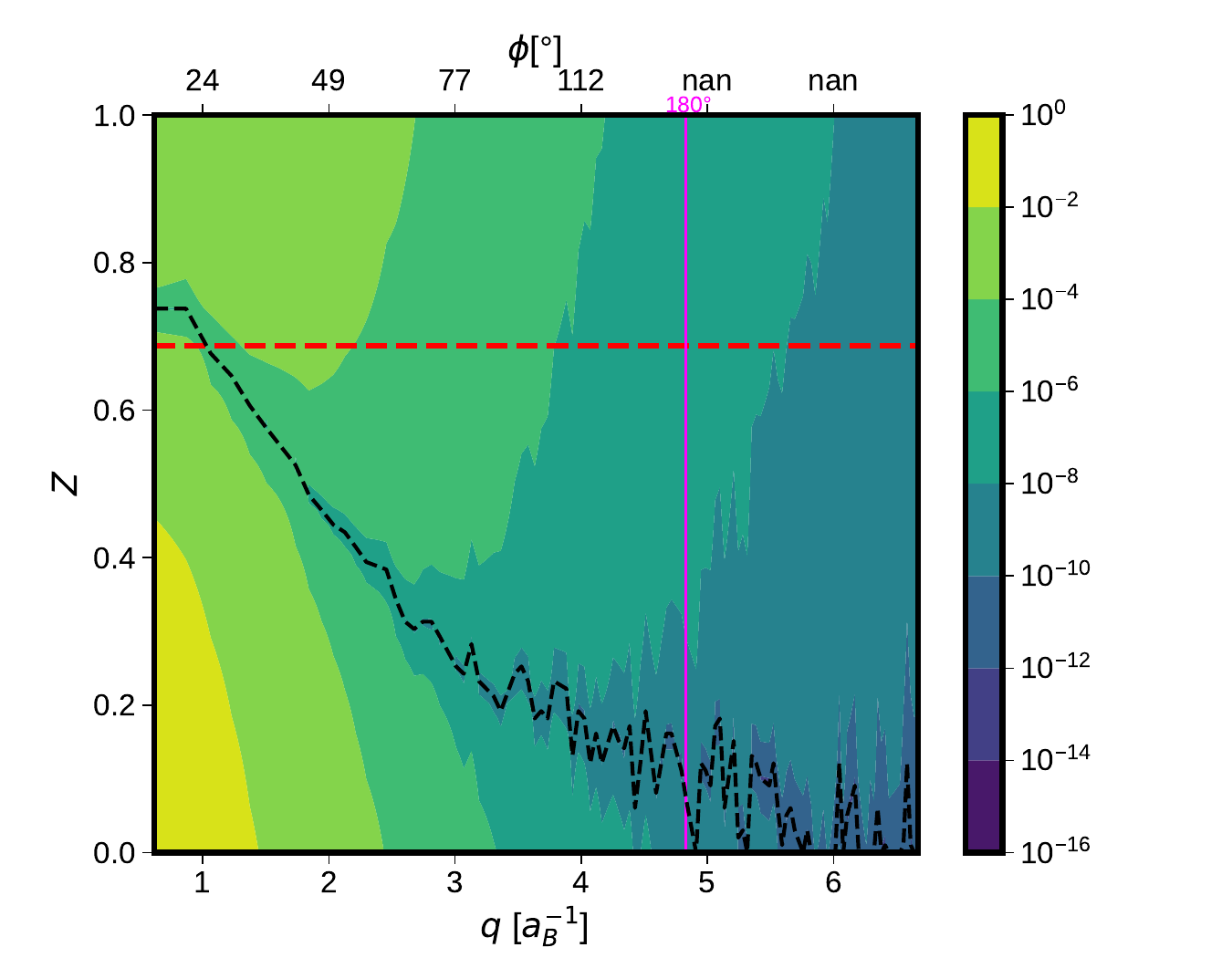}
    \includegraphics[width=0.48\linewidth]{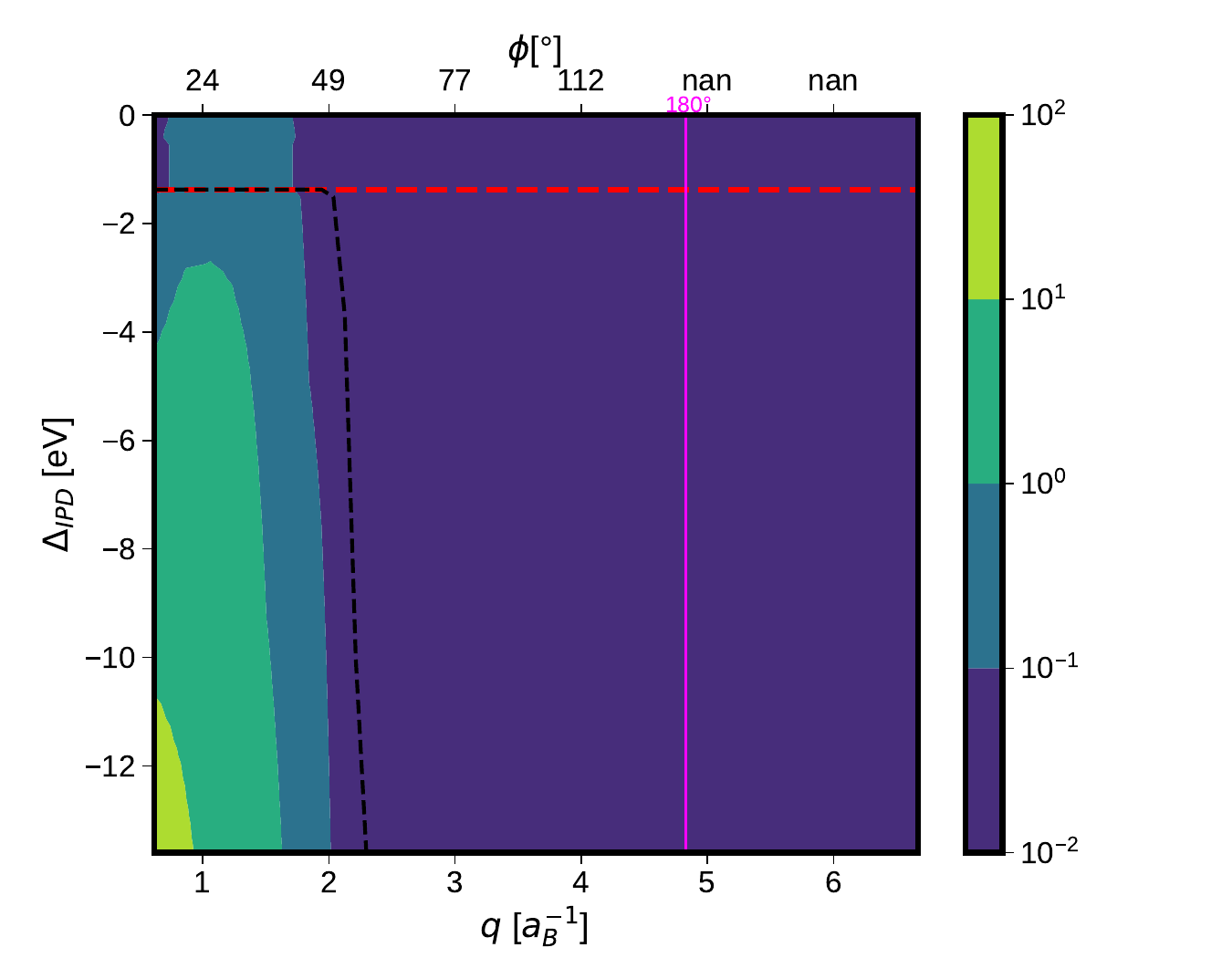}
    \includegraphics[width=0.48\linewidth]{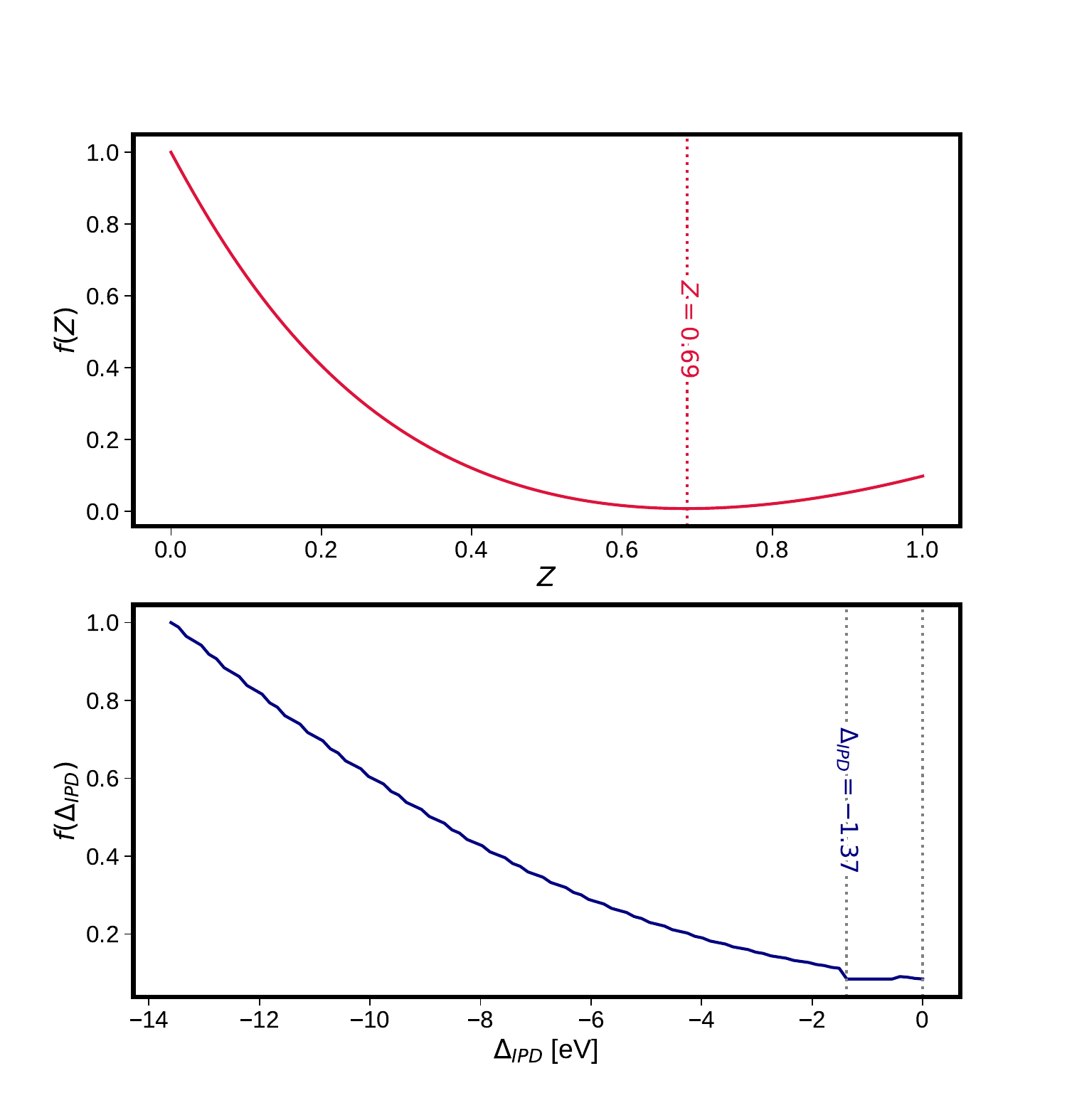}
    \includegraphics[width=0.46\linewidth]{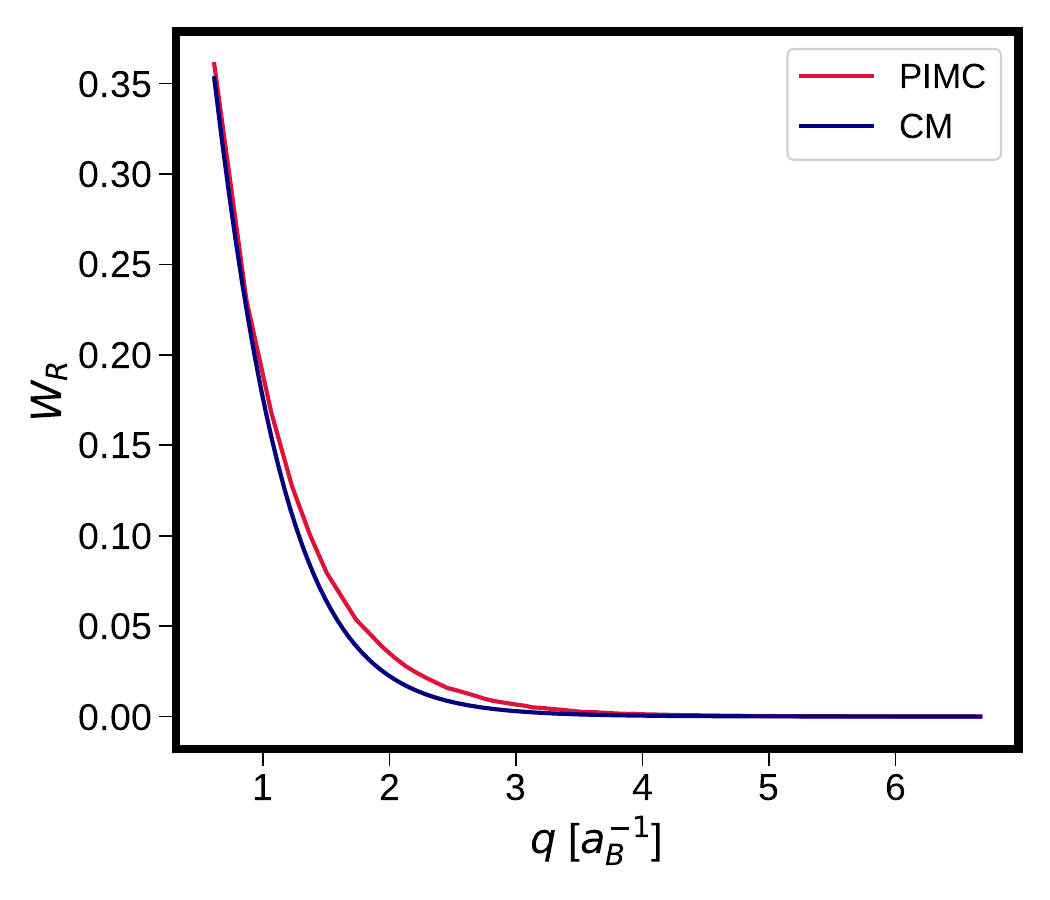}
    \caption{Results $N=32$, $r_s=2$, $\theta=1$ dataset. 
    Top left: $\chi$-error plot for the Rayleigh weight over $q-Z$ parameter space.
    Top right: $\chi$-error plot for the inelastic component over $q-\Delta_{\text{IPD}}$ parameter space.
    For each surface plot, corresponding scattering angles $\phi$ for a beam energy of $9\unit{keV}$ are shown on the top x-axis.
    The $180^{\circ}$ cut-off is shown in the solid pink horizontal line.
    Bottom left: ionization and IPD error, arbitrarily normalized to enforce $f(Z=0)=1$.
    Bottom right: best fit $W_R(q)$ plotted against the PIMC results.
    }
    \label{fig:N32-rs2-theta1}
\end{figure*}


\begin{figure*}[h]
    \includegraphics[width=0.48\linewidth]{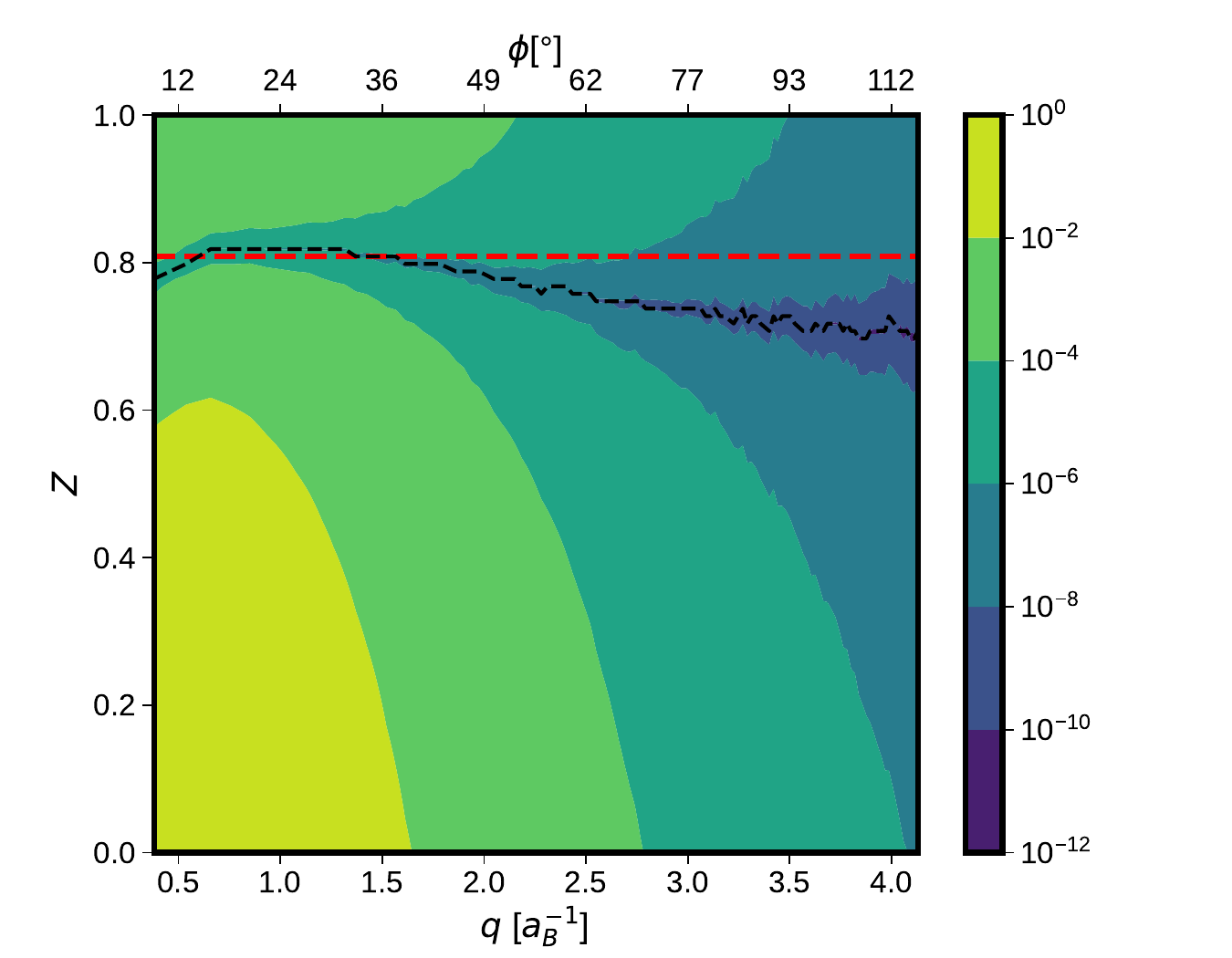}
    \includegraphics[width=0.48\linewidth]{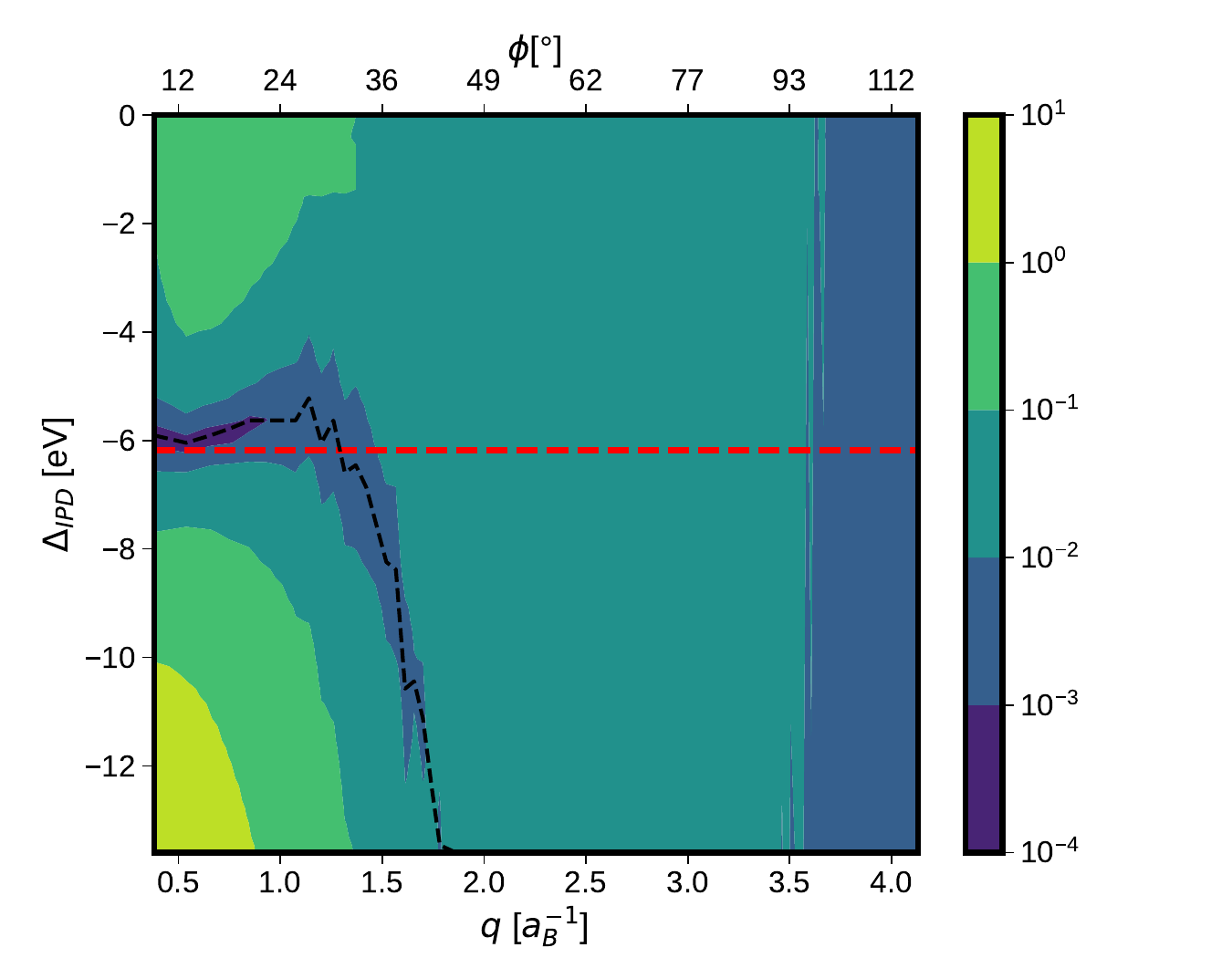}
    \includegraphics[width=0.48\linewidth]{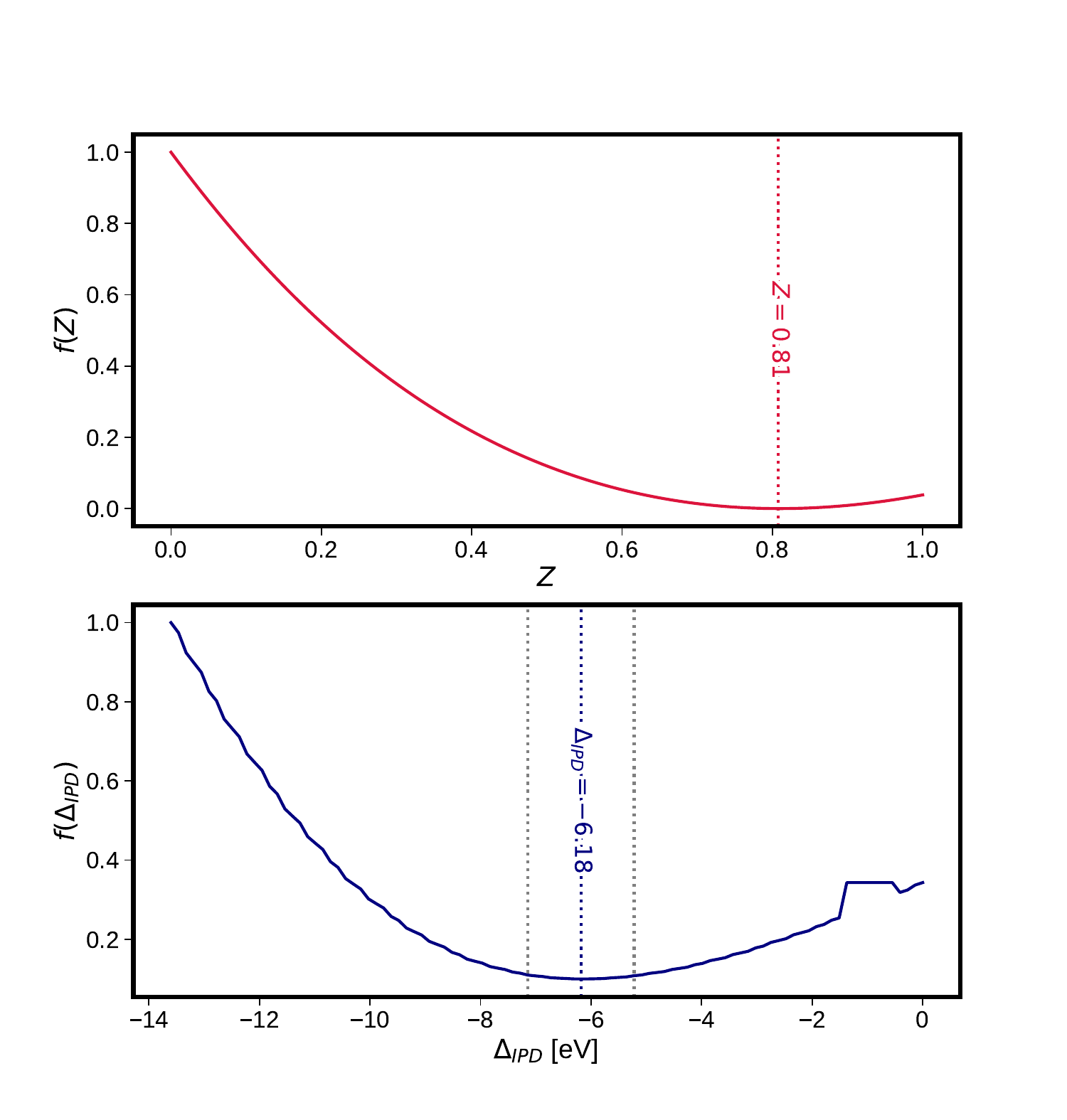}
    \includegraphics[width=0.46\linewidth]{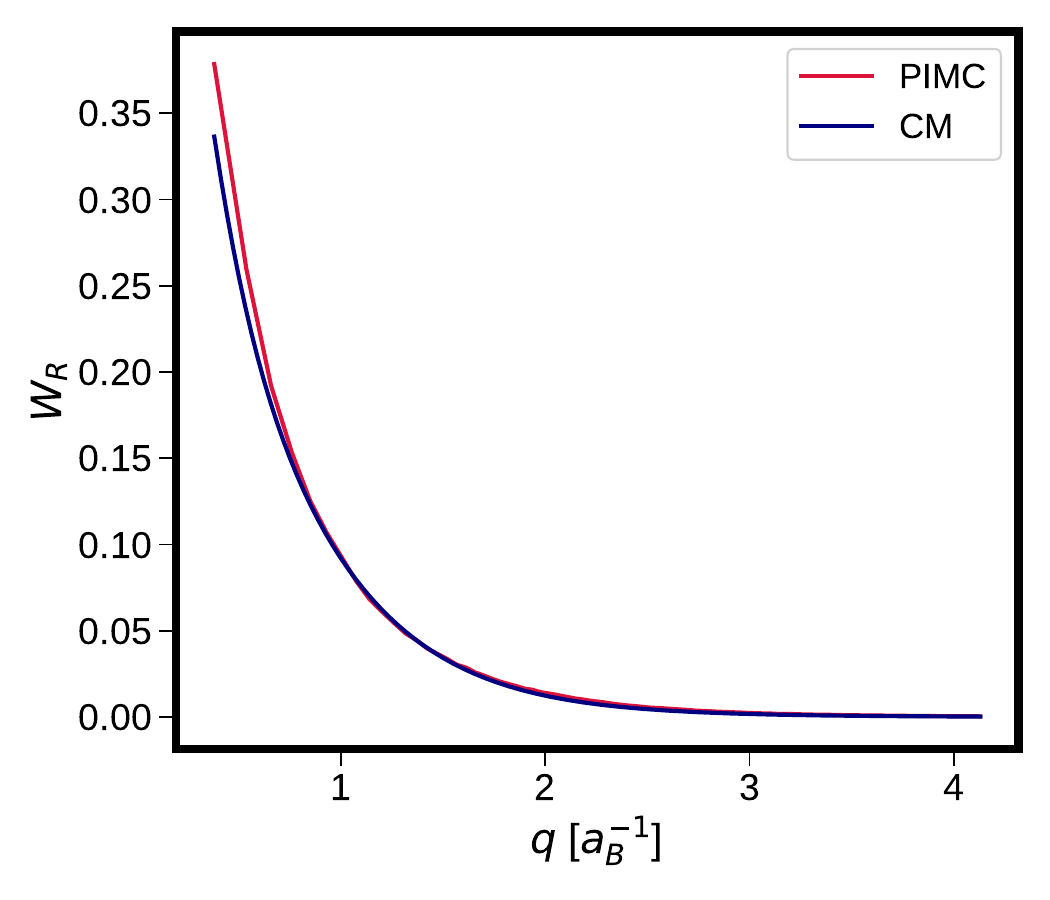}
    \caption{Results $N=32$, $r_s=3.23$, $\theta=2$ dataset. 
    Top left: $\chi$-error plot for the Rayleigh weight over $q-Z$ parameter space.
    Top right: $\chi$-error plot for the inelastic component over $q-\Delta_{\text{IPD}}$ parameter space.
    Bottom left: ionization and IPD error, arbitrarily normalized to enforce $f(Z=0)=1$.
    Bottom right: best fit $W_R(q)$ plotted against the PIMC results.
    }
    \label{fig:N32-rs3.23-theta2}
\end{figure*}

\begin{figure*}[h]
    \includegraphics[width=0.48\linewidth]{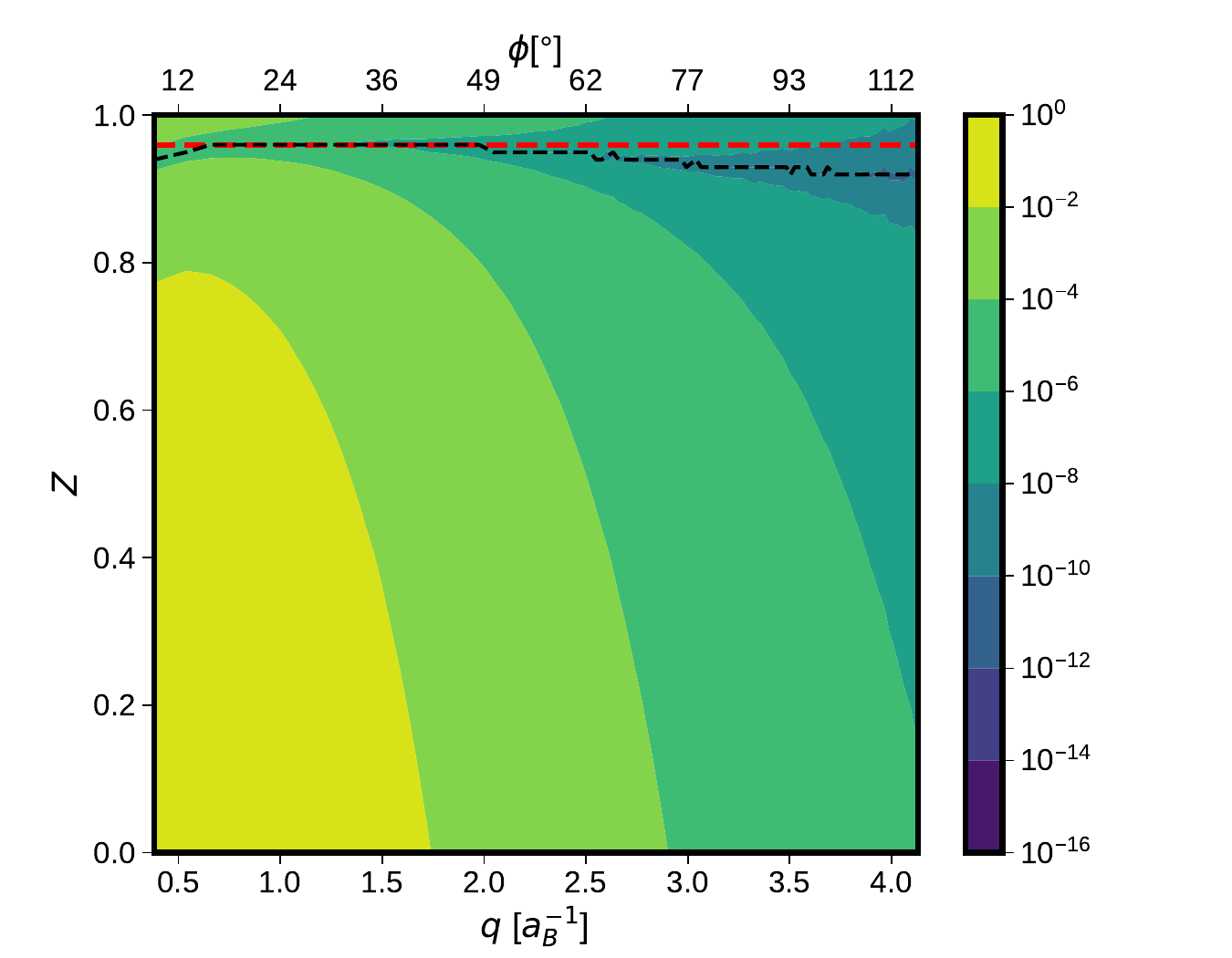}
    \includegraphics[width=0.48\linewidth]{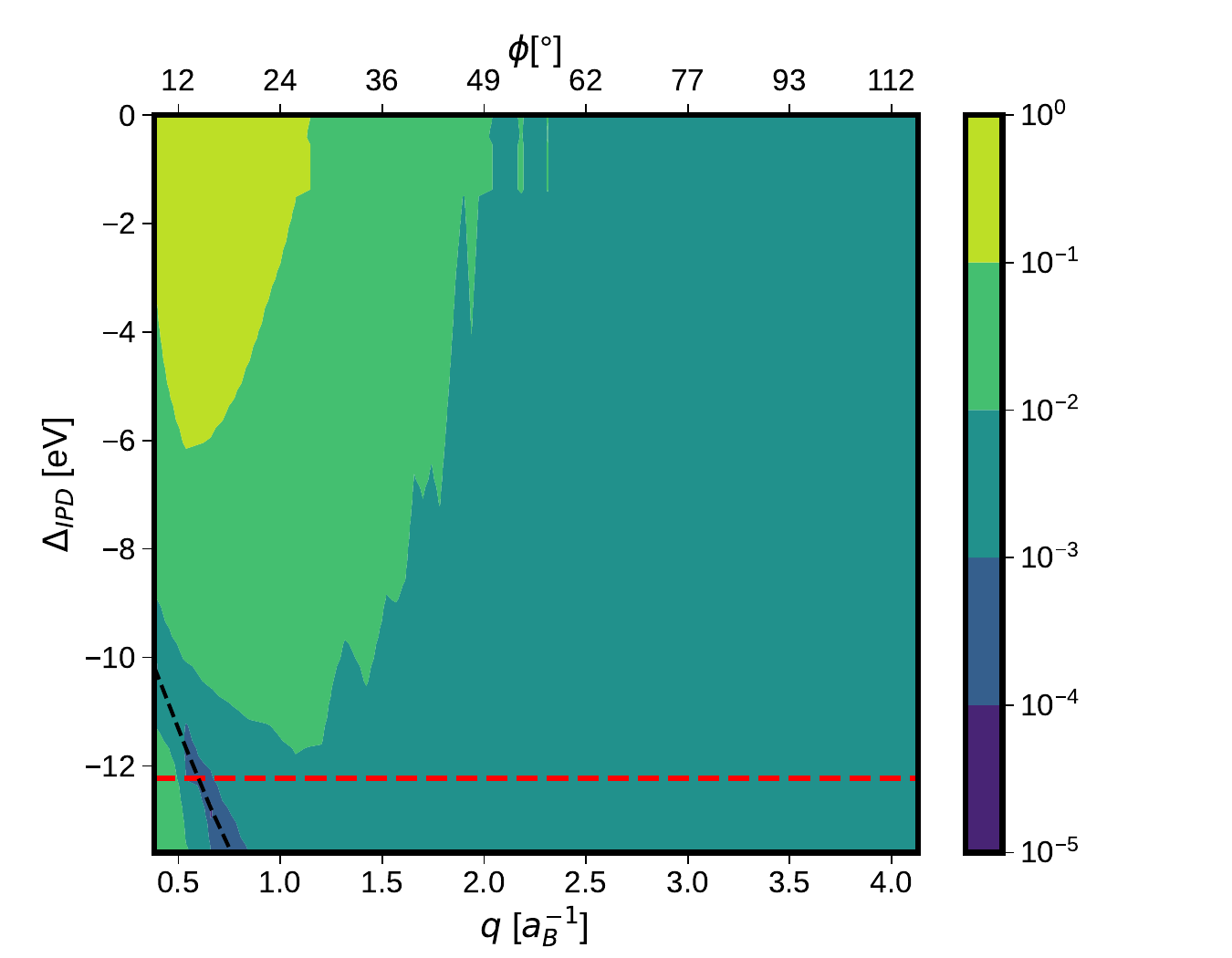}
    \includegraphics[width=0.48\linewidth]{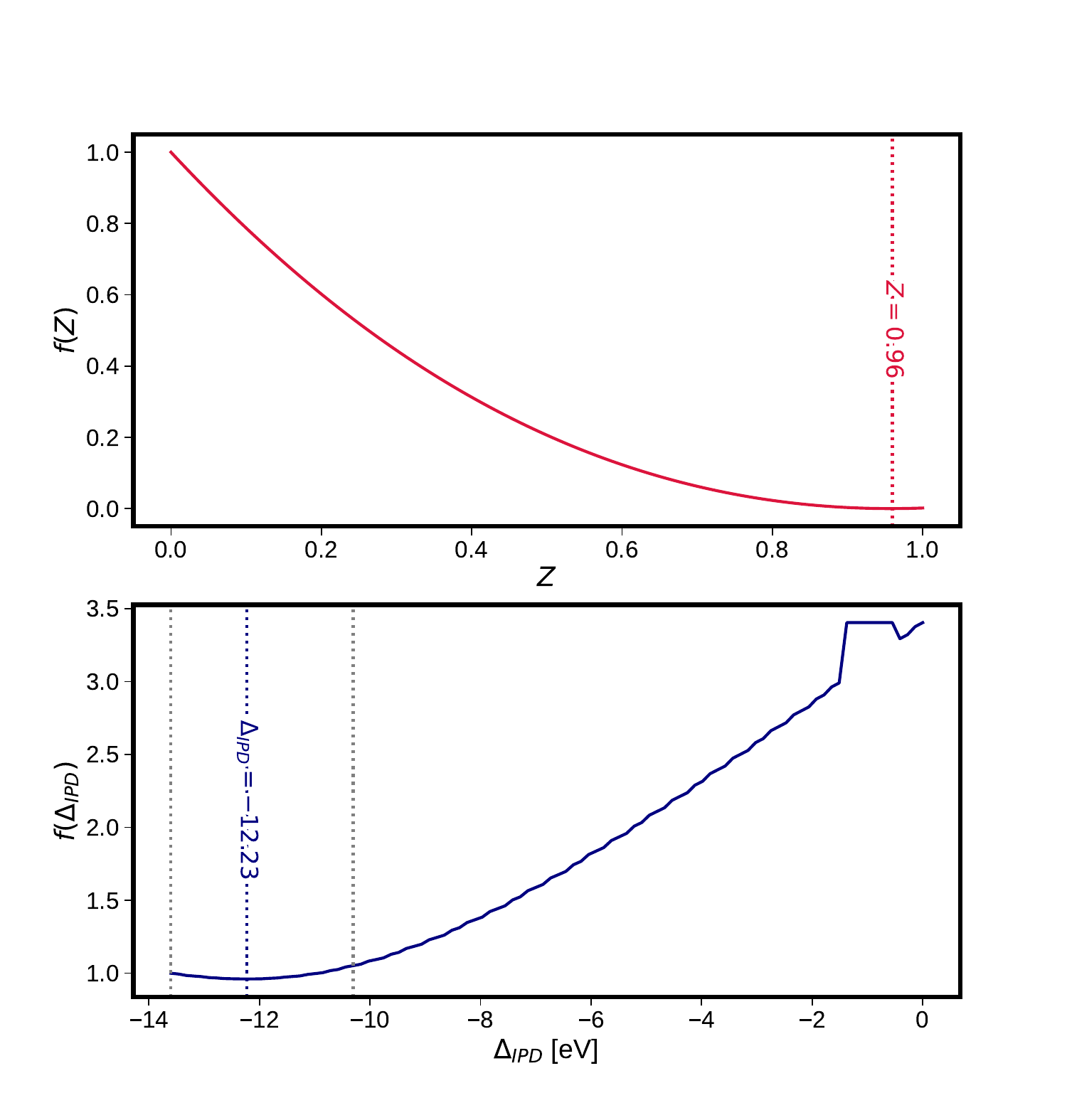}
    \includegraphics[width=0.46\linewidth]{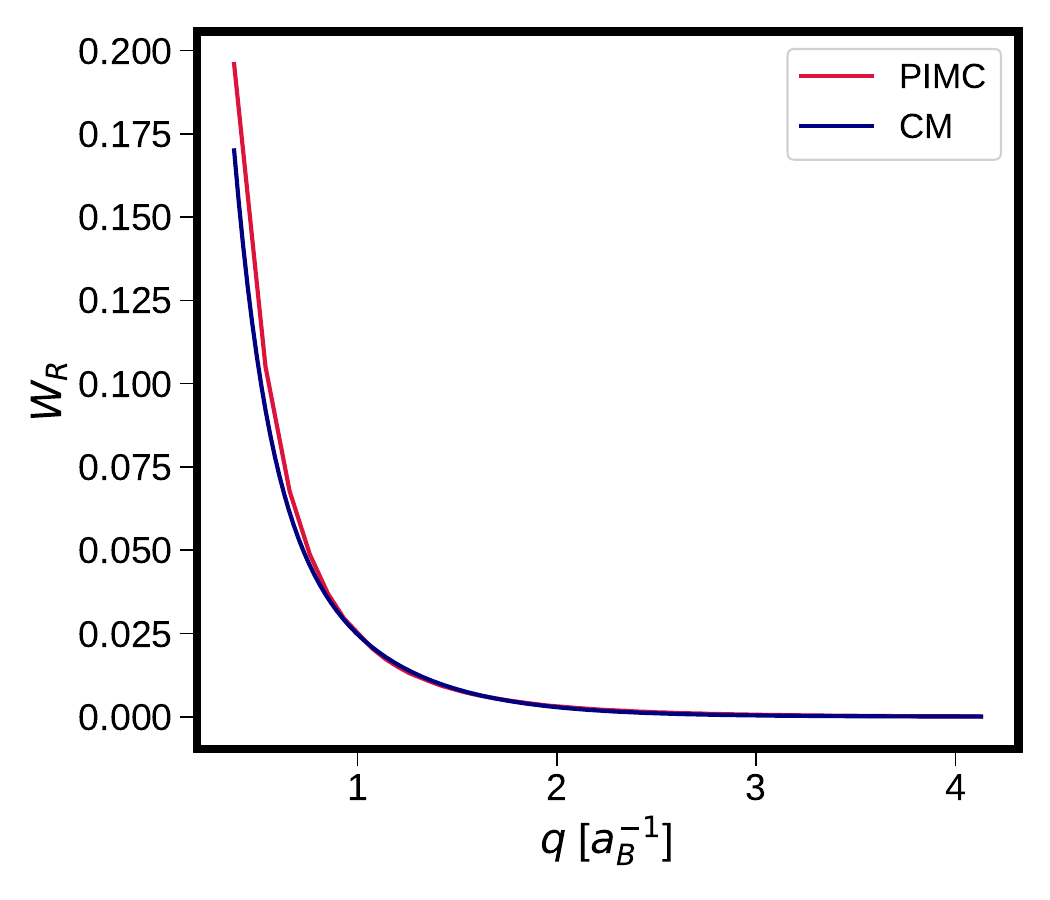}
    \caption{Results $N=32$, $r_s=3.23$, $\theta=4$ dataset. 
    Top left: $\chi$-error plot for the Rayleigh weight over $q-Z$ parameter space.
    Top right: $\chi$-error plot for the inelastic component over $q-\Delta_{\text{IPD}}$ parameter space.
    For each surface plot, corresponding scattering angles $\phi$ for a beam energy of $9\unit{keV}$ are shown on the top x-axis.
    Bottom left: ionization and IPD error, arbitrarily normalized to enforce $f(Z=0)=1$.
    Bottom right: best fit $W_R(q)$ plotted against the PIMC results.
    }
    \label{fig:N32-rs3.23-theta4}
\end{figure*}

\begin{figure*}[h]
    \includegraphics[width=0.48\linewidth]{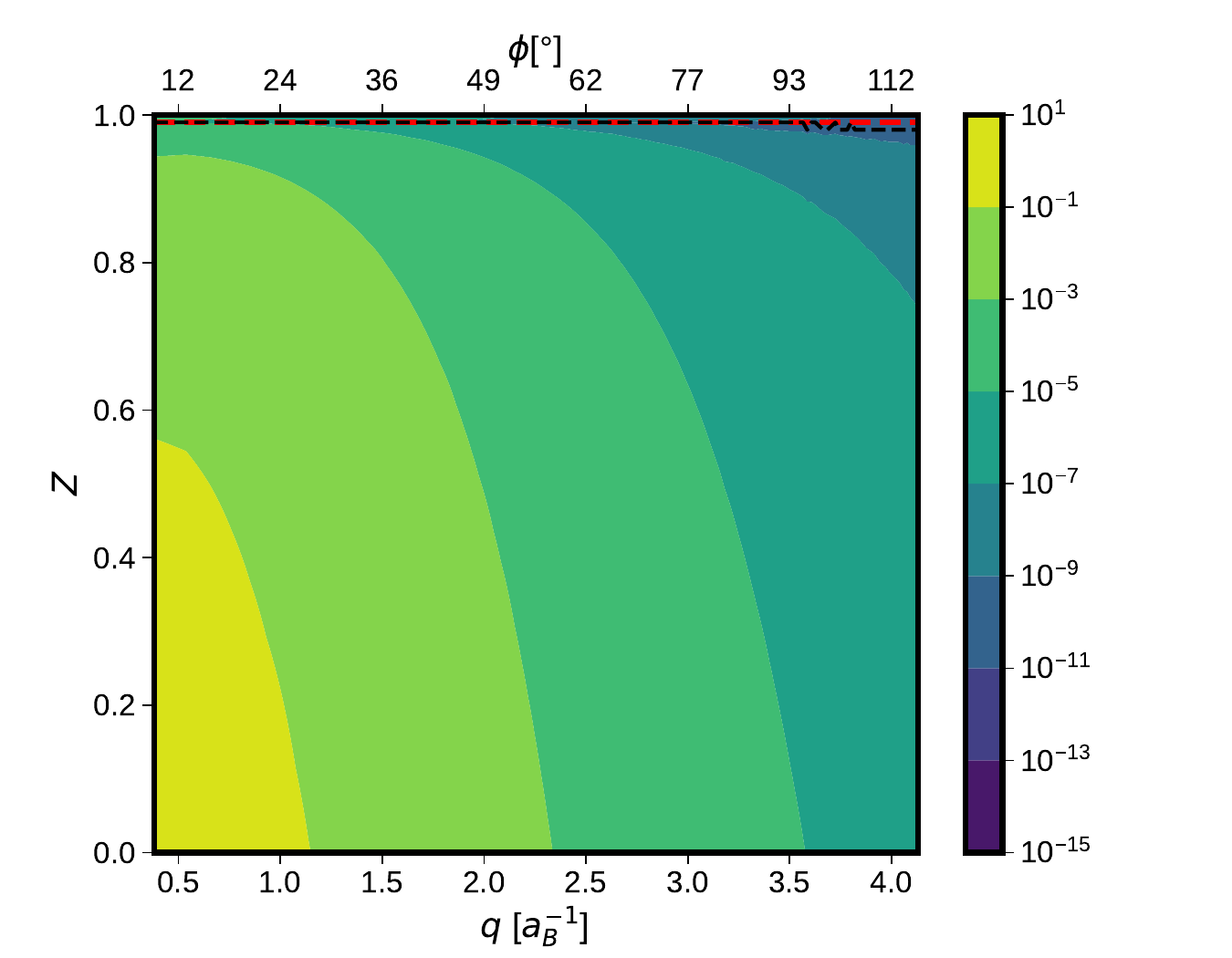}
    \includegraphics[width=0.48\linewidth]{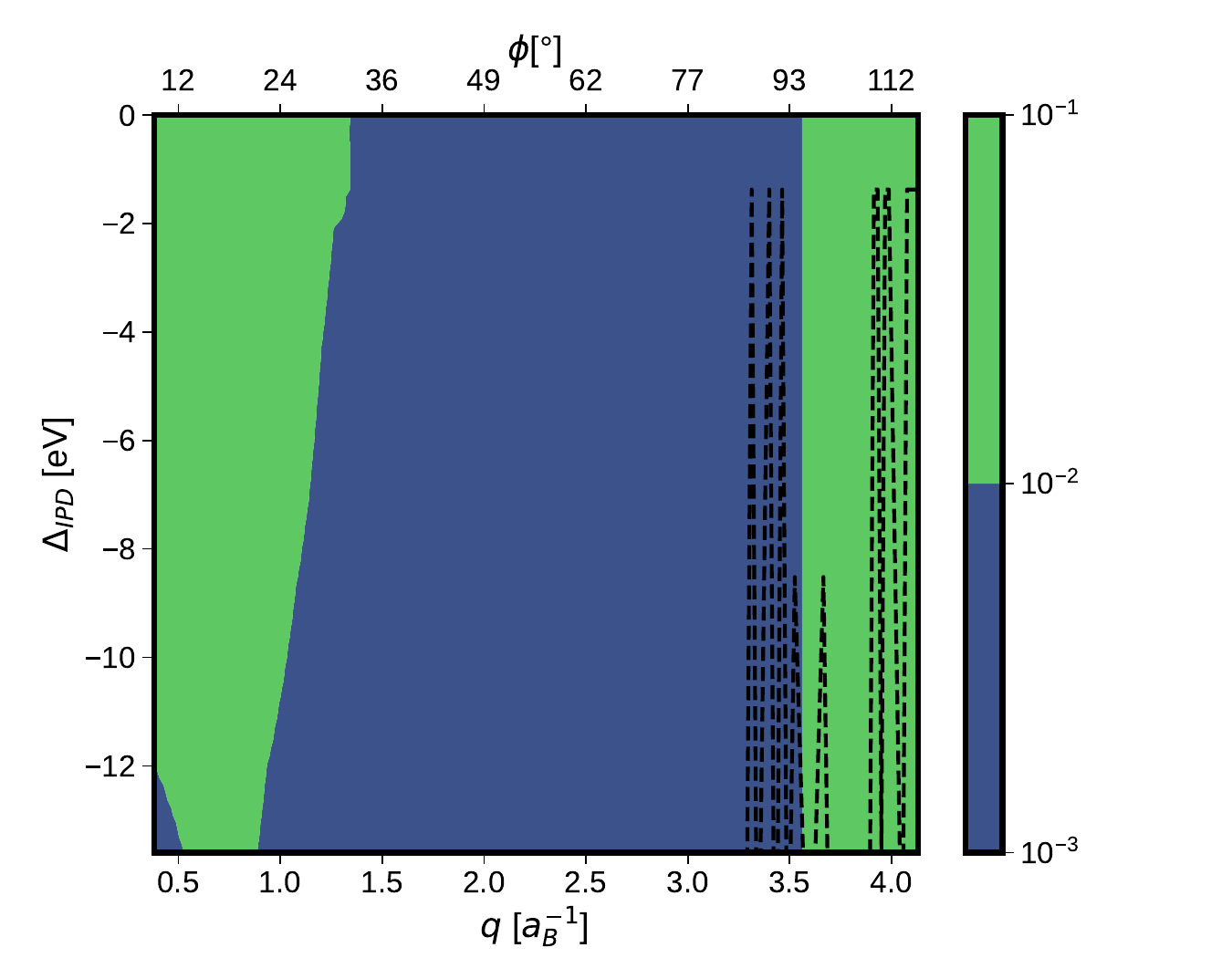}
    \includegraphics[width=0.48\linewidth]{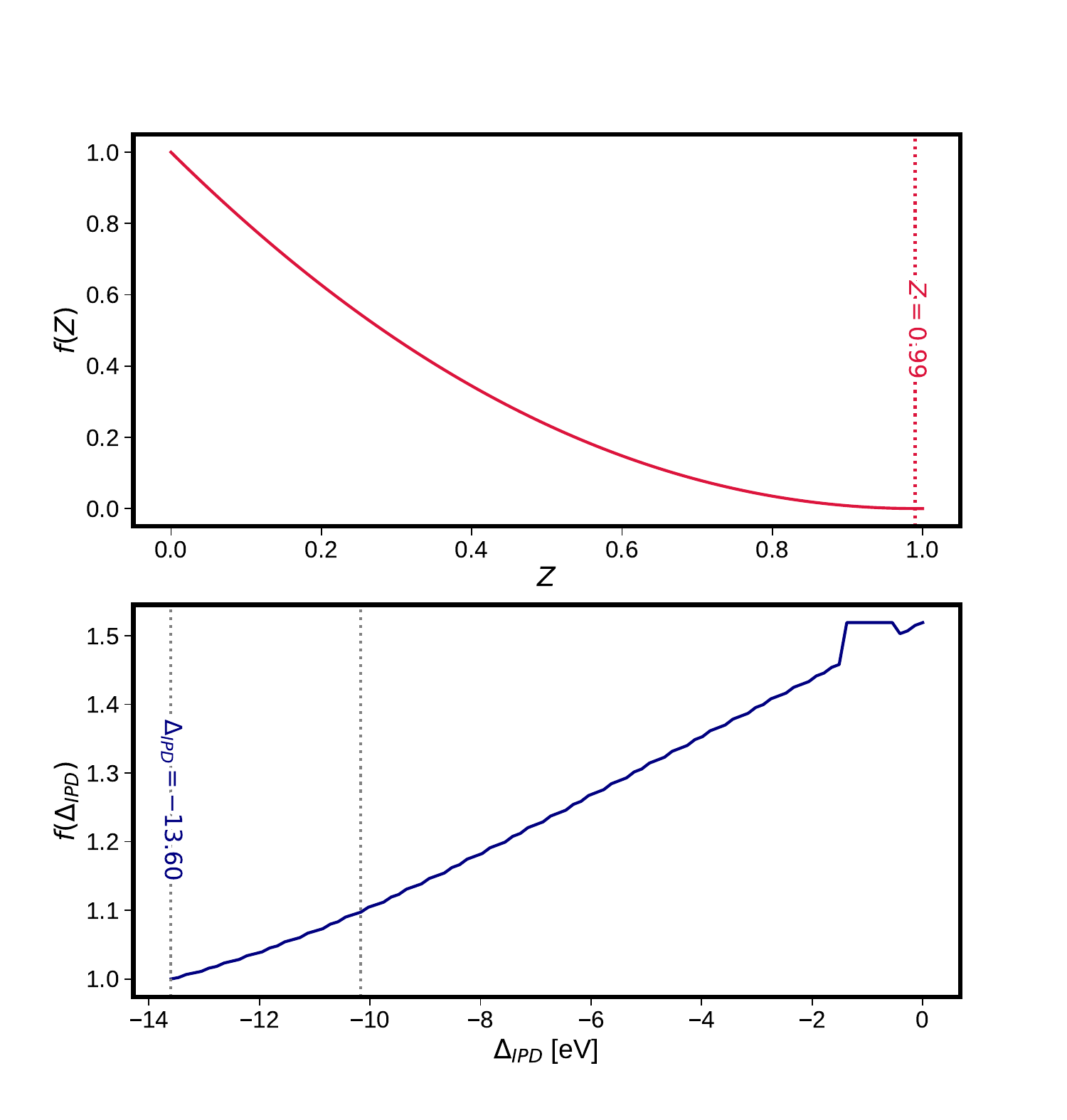}
    \includegraphics[width=0.46\linewidth]{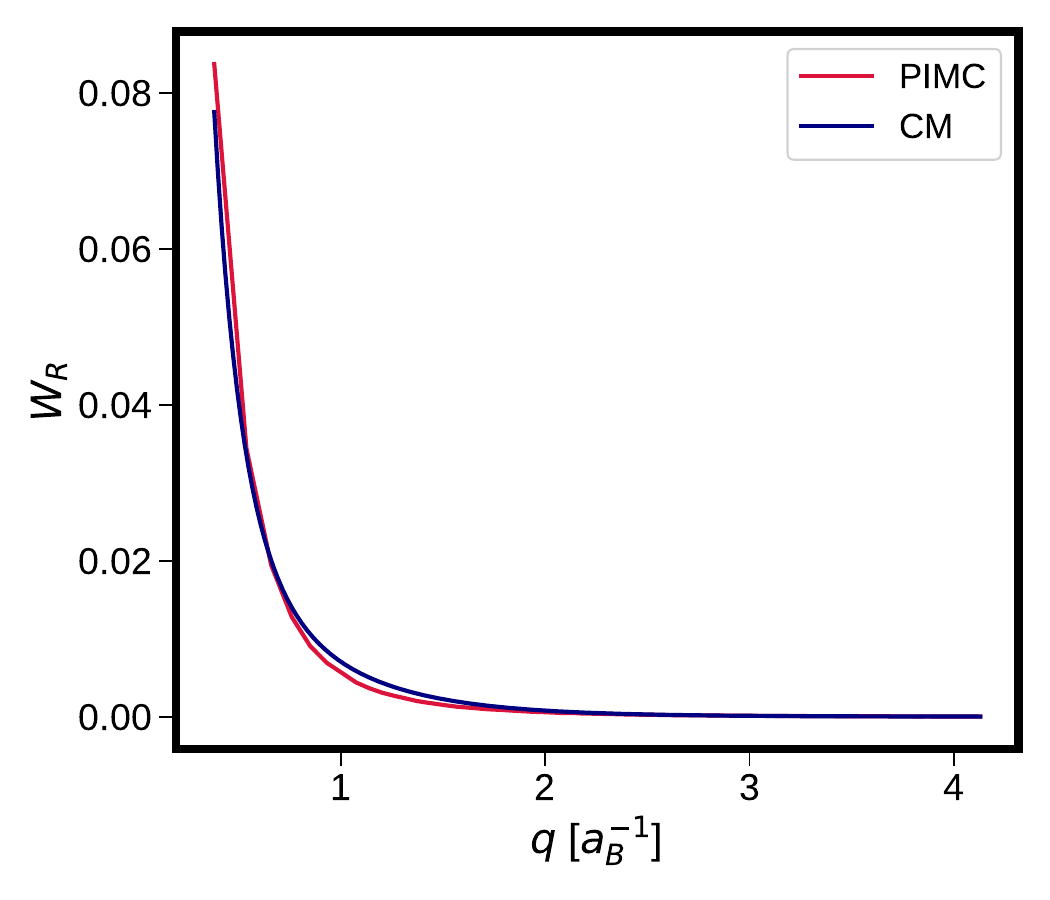}
    \caption{Results $N=32$, $r_s=3.23$, $\theta=8$ dataset. 
    Top left: $\chi$-error plot for the Rayleigh weight over $q-Z$ parameter space.
    Top right: $\chi$-error plot for the inelastic component over $q-\Delta_{\text{IPD}}$ parameter space.
    For each surface plot, corresponding scattering angles $\phi$ for a beam energy of $9\unit{keV}$ are shown on the top x-axis.
    Bottom left: ionization and IPD error, arbitrarily normalized to enforce $f(Z=0)=1$.
    Bottom right: best fit $W_R(q)$ plotted against the PIMC results.
    }
    \label{fig:N32-rs3.23-theta8}
\end{figure*}

\end{document}